\pdfoutput=1

\documentclass[11pt,a4paper]{article}

\usepackage{jheppub}
\usepackage{amsmath}
\usepackage{multicol,bbm}
\usepackage{enumerate}
\usepackage{bibentry}

\usepackage{amssymb}
\usepackage{bm}
\usepackage{multirow}
\usepackage{caption}
\usepackage{tabularx}
\usepackage{enumitem}
\usepackage{extarrows}
\usepackage{tikz}
\usepackage{verbatim} 
\usepackage{float} 

\graphicspath{{figures/}}  

\def\be{\begin{equation}}
\def\ee{\end{equation}}
\def\ba{\begin{eqnarray}}
\def\ea{\end{eqnarray}}

\def\CP1{\mathbb{CP}^1}
\def\SL2C{\mathrm{SL}(2,\mathbb{C})}

\def\Z2{\mathbb{Z}_2}

\def\su2{{SU(2)}}

\def\[{\left[}
\def\]{\right]}

\def\({\left(}
\def\){\right)}
\def\[{\left[}
\def\]{\right]}

\def\<{\langle}
\def\>{\rangle}

\def\i2{\frac{i}{2}}

\def\2F1{\,_2{\rm F}_1}

\begin{document}


\title{The momentum amplituhedron of SYM and ABJM from twistor-string maps}


\date{\today}
\author[a,b,c,d,e]{Song He,}
\author[f]{Chia{-}Kai Kuo,} 
 \author[a,d]{Yao{-}Qi Zhang}%
\affiliation[a]{CAS Key Laboratory of Theoretical Physics, Institute of Theoretical Physics, Chinese Academy of Sciences, Beijing 100190, China}
\affiliation[b]{
School of Fundamental Physics and Mathematical Sciences, Hangzhou Institute for Advanced Study, UCAS, Hangzhou 310024, China}
\affiliation[c]{ICTP{-}AP
International Centre for Theoretical Physics Asia{-}Pacific, Beijing/Hangzhou, China}
\affiliation[d]{School of Physical Sciences, University of Chinese Academy of Sciences, No.19A Yuquan Road, Beijing 100049, China}
\affiliation[e]{Peng Huanwu Center for Fundamental Theory, Hefei, Anhui 230026, P. R. China}
\affiliation[f]{Department of Physics and Center for Theoretical Physics, National Taiwan University, Taipei 10617, Taiwan}

\emailAdd{songhe@itp.ac.cn}
\emailAdd{chiakaikuo@gmail.com}
\emailAdd{zhangyaoqi@itp.ac.cn}

\date{\today}

\abstract{We study remarkable connections between twistor-string formulas for tree amplitudes in ${\cal N}=4$ SYM and ${\cal N}=6$ ABJM, and the corresponding momentum amplituhedron in the kinematic space of $D=4$ and $D=3$, respectively. Based on the Veronese map to positive Grassmannians, we define a twistor-string map from $G_{+}(2,n)$ to a $(2n{-}4)${-}dimensional subspace of the 4d kinematic space where the momentum amplituhedron of SYM lives. We provide strong evidence that the twistor-string map is a diffeomorphism from $G_+(2,n)$ to the interior of momentum amplituhedron; the canonical form of the latter, which is known to give tree amplitudes of SYM, can be obtained as pushforward of that of former. We then move to three dimensions: based on Veronese map to orthogonal positive Grassmannian, we propose a similar twistor-string map from the moduli space ${\cal M}_{0,n}^+$ to a $(n{-}3)${-}dimensional subspace of 3d kinematic space. The image gives a new positive geometry which conjecturally serves as the momentum amplituhedron for ABJM; its canonical form gives the tree amplitude with reduced supersymmetries in the theory. We also show how boundaries of compactified ${\cal M}_{0,n}^+$ map to boundaries of momentum amplituhedra for SYM and ABJM corresponding to factorization channels of amplitudes, and in particular for ABJM case the map beautifully excludes all unwanted channels.}


\maketitle

\section{Introduction}\label{sec1}

The complete tree-level S-matrix of ${\cal N}=4$ supersymmetric Yang-Mills and ${\cal N}=8$ supergravity in four dimensions can be presented as integrals over the moduli space of maps from the n-punctured sphere to twistor space as well as momentum space~\cite{Witten:2003nn, PhysRevD.70.026009,Cachazo:2012da,Cachazo:2012kg,Cachazo:2012pz}. These formulas are precursors of Cachazo-He-Yuan (CHY) formulas for tree-level S-matrix of gluons and gravitons in any dimension~\cite{Cachazo:2013gna, Cachazo:2013hca}: historically it was by considering a modified version of these formulas in $D=4$ where amplitudes are dressed with wave functions imposing on-shell conditions~\cite{Cachazo:2013iaa}, that has led to the discovery of scattering equations~\cite{Cachazo:2013gna} and CHY formulas in general dimensions~\cite{Cachazo:2013hca,Cachazo:2013iea,Cachazo:2014nsa, Cachazo:2014xea}. By specifying CHY formulas to specific dimensions, one obtains these twistor-string-like formulas not only in $D=4$ but also in  $D=3,6$ {\it etc.}, which utilize spinor-helicity variables and allow for natural supersymmetric extensions. These formulas apply to amplitudes in super-Yang-Mills (SYM), super-gravity (SUGRA) and supersymmetric Chern-Simons such as Aharony-Bergman-Jafferis-Maldacena (ABJM) and Bagger-Lambert-Gustavsson (BLG) theories in $D=3$~\cite{Cachazo:2013iaa}, as well as a wide class of amplitudes in $D=6$, including SYM, SUGRA and $D_5$- and $M_5$-brane amplitudes~\cite{Cachazo:2018hqa, Schwarz:2019aat, Geyer:2018xgb}.

A seemingly unrelated line of research concerns {\it positive geometries}~\cite{Arkani-Hamed:2017tmz} underlying scattering amplitudes. Based on the dual formulation of planar ${\cal N}=4$ SYM in terms of positive Grassmannian~\cite{Arkani-Hamed:2012zlh}, a geometric object called the amplituhedron was proposed~\cite{Arkani-Hamed:2013jha, Arkani-Hamed:2013kca}, whose {\it canonical forms}~\cite{Arkani-Hamed:2017tmz} encode tree amplitudes and all-loop integrands of the theory; it was defined as the image of positive Grassmannian via a map that was given in terms of bosonized momentum twistor variables~\cite{Hodges:2009hk}. More recently, a new geometric object of dimension $2n{-}4$, dubbed {\it momentum amplituhedron} was proposed in a similar fashion but directly in momentum space (spanned by spinors $(\lambda, \tilde\lambda)$)~\cite{Damgaard:2019ztj}; its canonical form encodes $n$-point tree amplitudes of ${\cal N}=4$ SYM, which was a nice $d\log$ differential form first considered in~\cite{He:2018okq} (such forms can be used for unifying all helicity amplitudes in general theories at least in $D=4$). This has provided a link between these two lines of research: the $2n{-}4$ form can be obtained as a pushforward of canonical form of the moduli space of twistor strings, or $G_+(2,n)$, by summing over all solutions of scattering equations in $D=4$~\cite{Cachazo:2013iaa}. Thus the momentum amplituhedron must be directly related to twistor-string theory and scattering equations in $D=4$, which then must allow us to explore the geometric origin of the latter. In this paper we come back to the idea of~\cite{He:2018okq} and revisit the momentum amplituhedron of SYM as the image of $G_+(2,n)$ via a (one-to-one) map defined by twistor-string formulas, or rather the $D=4$ scattering equations. 

On the other hand, while we have twistor-string-like formulas also in $D=3,6$ for various theories~\cite{Huang:2012vt}, so far no positive geometries have been obtained for amplitudes in these theories. Inspired by the twistor-string maps in $D=4$, we will propose a similar map in $D=3$, where the moduli space is  ${\cal M}_{0,n}^+$ since the ``$t$ variables" of $G_+(2,n)$ are completely determined~\cite{Cachazo:2013iaa}. We will show that the image of our map gives a $(n{-}3)$-dimensional positive geometry which can serve as the {\it ABJM momentum amplituhedron} in $D=3$ kinematic space; the latter can be defined in a similar fashion as SYM in~\cite{Damgaard:2019ztj} without referring to the twistor-string map, and its canonical forms give ABJM amplitudes with reduced SUSY~\footnote{The authors of~\cite{Huang:toa} have independently discovered ABJM momentum amplituhedron with a definition similar to the SYM case in $D=4$. It has some overlap with sec.~\ref{sec:3d} of this paper.}. 

\paragraph{Kinematic associahedron from scattering-equation map} 
Before proceeding, let us quickly review the parallel but simpler story for the tree-level amplitudes in bi-adjoint $\phi^3$ theory in general $D$~\cite{Arkani-Hamed:2017mur}, which has provided an example of our map and pushforward. The tree-level amplituhedron of bi-adjoint $\phi^3$ theory, as directly formulated in the kinematic space of general $D$, is the so-called ABHY associahedron~\cite{Arkani-Hamed:2017mur}. Note that the kinematic space can be spanned by the so-called planar variables $X_{i,j}$ for $1\leq i<j-1<n$ (with dimension $n\left(n{-}3\right)/2$), which are identified with the diagonals of an n-gon with edges given by $p_{1},p_{2},\cdots,p_{n}$, $X_{i,j}=\left(\sum_{a=i}^{j-1}p_{a}\right)^{2}:=s_{i,i+1,\cdots j-1}$. The kinematic associahedron, $\mathcal{A}\left(1,\cdots,n\right)=\mathcal{A}_{n-3}$ is a $\left(n-3\right)$-dim polytope defined as the intersection of the {\it positive region} 
$\Delta_{n}=\left\{ X_{i,j}\geq0\ \mathrm{for\,all}\ 1\leq i<j-1<n\right\}$ with a {\it subspace} of dimension $\left(n-3\right)$ defined by the following $\left(n-2\right)\left(n-3\right)/2$ conditions:\begin{equation}\label{eq:ABHY}\begin{split}H(1,2,\cdots,n):=\{C_{i,j}=X_{i,j}+X_{i+1, j+1}-X_{i,j+1}-X_{i+1,j}\\{\rm are\ positive\ constants, \, for}\ 1\leq i<j{-}1<n{-}1\} \end{split}\end{equation}
We have $\mathcal{A}_{n-3}=\Delta_{n}\cap H\left(1,2,\cdots,n\right)$ which gives the $(n{-}3)$-dim associahedron with each facet (vertex) representing a planar pole (planar cubic tree diagram), and its canonical form gives the planar $\phi^3$ tree amplitude: \begin{equation}	\Omega_{A_{n-3}}
=\prod_{a=1}^{n-3}dX_{i_{a},j_{a}}M\left(1,2,\cdots,n\right)\end{equation}
where $M\left(1,2,\cdots,n\right):=m\left(1,2,\cdots,n|1,2,\cdots,n\right)$ is the “diagonal” bi-adjoint amplitudes, or the sum of all planar cubic trees. 

Very nicely, the ABHY associahedron turns out to be the image of the {\it positive moduli space}, ${\cal M}_{0,n}^+$, via a map originated from the CHY scattering equations~\cite{Cachazo:2013hca}:
\begin{equation}
E_{i}:=\sum_{j=1 ; j \neq i}^{n} \frac{s_{i j}}{\sigma_{i, j}}=0 \quad \text { for } i=1, \ldots, n
\end{equation}
By pulling back these equations on the subspace $H(1,2,\cdots, n)$, we find that the resulting {\it scattering-equation map} a diffeomorphism from ${\cal M}_{0,n}^+$ to the interior of ${\cal A}_{n{-}3}$~\cite{Arkani-Hamed:2017mur}. Note that $\mathcal{M}_{0,n}^{+}$ can be constructed as the positive Grassmannian $G_{>0}\left(2,n\right)$ modded out by the torus action $\mathbb{R}_{>0}^{n}$, {\it i.e.}  $2\times n$ matrix with $\sigma_{i}$ for each puncture and $\sigma_{i}<\sigma_{j}$ for $i<j$:

$$\left(\begin{array}{ccccc}1 & 1 & \cdots & 1 & 1\\\sigma_{1} & \sigma_{2} & \cdots & \sigma_{n-1} & \sigma_{n}\end{array}\right). $$

The well-known compacitification of $\mathcal{M}_{0,n}^{+}$ produces the boundary structure of the associahedron known as the worldsheet associahedron, which can be most conveniently described by $u$ variables~\cite{Arkani-Hamed:2017mur, Arkani-Hamed:2019mrd, Arkani-Hamed:2019plo} as we will review later. Its canonical form is given by\begin{equation}   \mathbf{\omega}_{n}^{\mathrm{WS}}=\frac{1}{\mathrm{vol}\,\mathrm{SL}\left(2\right)}\prod_{a=1}^{n}\frac{d\sigma_{a}}{\sigma_{a}-\sigma_{a+1}}=\frac{1}{\mathrm{vol}\,\mathrm{SL}\left(2\right)\times\mathrm{GL}\left(1\right)^{n}}\prod_{a=1}^{n}\frac{d^{2}C_{a}}{\left(a\,a+1\right)}.\end{equation}
Given the diffeomorphism, one can compute the canonical form $\Omega({\cal A}_{n{-}3})$ as the {\it pushforward} of $\mathbf{\omega}^{\rm WS}_n$, by summing over $(n{-}3)!$ solutions of the scattering equations:
\be
\mathbf{\Omega}({\cal A}_{n{-}3})=\sum_{\rm sol.}^{(n{-}3)!} \mathbf{\omega}^{\rm WS}_n
\ee
which is equivalent to the original CHY formula for bi-adjoint $\phi^3$ amplitudes, including all the double-partial amplitudes, $m(\alpha|\beta)$, obtained by pulling $\alpha$-ordered form $\Omega(\alpha)$ back to subspaces corresponding to the $\beta$ ordering~\cite{Arkani-Hamed:2017mur}.  

\paragraph{Outline of the paper} What we will study in this paper is a parallel picture for momentum amplituhedron of ${\cal N}=4$ SYM~\cite{Damgaard:2019ztj}, as well as a new positive geometry which we conjecture to be the momentum amplituhedron of ABJM. Although much more involved than kinematic associahedron, both (non-polytopal) geometries can be defined similarly as the intersection of a top-dimensional region in the kinematic space of $D=4$ (or $D=3$), which requires only positivity but also certain ``winding" conditions~\cite{Damgaard:2020eox}, and a subspace of dimension $2n{-}4$ (or $n{-}3$), as we will discuss in order. Again similar to associahedron, we will show that these geometries are images of their moduli space via certain maps in $D=4$ and $D=3$ respectively. In section~\ref{4d}, we will propose that the SYM momentum amplituhedron is nothing but the image of $G_+(2,n)$ via a {\it twistor-string map} defined by the pullback of $D=4$ scattering equations; in particular we provide strong evidence that the map is in fact a diffeomorphism, which then implies that one can pushforward the canonical form of $G_+(2,n)$ to obtain that of the momentum amplituhedron for tree amplitudes in ${\cal N}=4$ SYM as expected. We then conjecture in section~\ref{sec:3d} that the direct analog in $D=3$, namely the image of ${\cal M}_{0,n}^+$ via a $D=3$ twistor-string map, gives the momentum amplituhedron for ABJM; the pushforward then gives the tree amplitudes of ABJM reduced supersymmetries. We find it very satisfying that these three ``kinematic amplituhedra" for general $D$, $D=4$ and $D=3$ cases can be ``unified" under the same theme: certain (one-to-one) maps from moduli space to kinematic space derived from scattering equations, and the pushforward of corresponding canonical forms. In particular, we will see that the $D=4$ and $D=3$ maps indeed know about different boundary structures of their momentum amplituhedra, which reflect different pole structures and factorizations of SYM and ABJM amplitudes. 
\paragraph{Scattering equations and twistor-string formulas in $D=4,3$} Next we will review scattering equations in $D=4,3$ and the corresponding twistor-string-like formulas. In $D=4$ and expressed in terms of spinor variables $(\lambda_i, \tilde\lambda_i)$ for $i=1,2,\cdots, n$, the CHY scattering equations fall into sectors labelled by $d=0, 1, \cdots, n{-}3$~\cite{Geyer:2014fka,He:2016vfi}, which denote the degree of polynomial of $\sigma_i$ for the spinors:
\begin{equation}\label{eq:polyrho}
\lambda_i=t_i \sum_{m=0}^d \rho_m \sigma_i^m 
\,,\quad
\tilde{\lambda}_i=\tilde{t}_i \sum_{\tilde{m}=0}^{\tilde{d}} \tilde{\rho}_{\tilde{m}} \sigma_i^{\tilde{m}}\,,
\end{equation}
where the degrees are $(d, \tilde{d}:=n{-}2{-}d)$ respectively, and $\tilde{t}_i=t_i^{-1} \prod_{j\neq i} \sigma^{-1}_{i,j}$. Note that the null momentum $p_i^{\alpha, \dot{\alpha}} =\lambda_i^\alpha \tilde{\lambda}_i^{\dot{\alpha}}$ is then given by a degree-$(n{-}2)$ polynomial, modulo the overall denominator $v_i:=\prod_{j\neq i} \sigma_{i,j}$~\cite{Cachazo:2013zc} (such that momentum conservation is automatically satisfied). The $t_i$'s and $\sigma_i$'s can be combined as $G(2,n)$, with $i$-th column given by $c_i:=t_i (1, \sigma_i)$ (note $d^2 c_i=t_i d t_i d\sigma_i$). We will thus call this $G(2,n)$ the moduli space of $D=4$ twistor-string formulas. Equivalently we can use $\tilde{t}_i$'s as part of the moduli space, which form another $G(2,n)$ related to the previous one by parity. 

It is natural to integrate out the $2(d{+}1)$ $\rho$'s and $2(\tilde{d}{+}1)$ $\tilde{\rho}$'s, which then yields constraints in terms of the following Veronese map: define $C_{\alpha, i}=t_i \sigma_i^{\alpha-1}$ for $\alpha=1,\cdots, k:=d{+}1$, and similarly $C^{\perp}_{\dot{\alpha}, i}=\tilde{t}_i \sigma_i^{\dot{\alpha}-1}$ for $\dot{\alpha}=1, \cdots, n{-}k=\tilde{d}{+}1$; note that $C \in G(k,n)$ and $C^{\perp} \in G(n{-}k, n)$, thus we have maps from $G(2,n)$ to $G(k,n)$ and $G(n{-}k, n)$, which are in the orthogonal complement of each other, $C \cdot C^{\perp}:=\sum_{i=1}^n C_{\alpha, i} C^{\perp}_{\tilde{\alpha}, i}=0$. Kinematically speaking, these are of course familiar already in the original story of Grassmannian formulas for tree amplitudes and leading singularities of SYM. For more details on positive Grassmannian and its BCFW cells for SYM, please refer to appendix~\ref{review}. 

The $D=4$ scattering equations now take the form that $C$ is orthogonal to $\tilde\lambda \in G(2,n)$ and $C^{\perp}$ is orthogonal to $\lambda \in G(2,n)$ (for $\alpha=1,\cdots, k$ and $\tilde{\alpha}=1,\cdots, n{-}k$): 
\begin{equation}\label{eq:cdl1}
    C \cdot \tilde\lambda\equiv \sum_{i=1}^n t_i \sigma^{\alpha-1}_i \tilde\lambda_i=0\,, \quad C^{\perp} \cdot \lambda\equiv\sum_{i=1}^n \tilde{t}_i \sigma^{\tilde{\alpha}-1}_i \lambda_i=0\,.
\end{equation}
As shown in~\cite{Cachazo:2013iaa}, there are Eulerian number of solutions, $E_{n{-}3,k{-}2}$, for $D=4$ scattering equations in sector labelled by $(k, n{-}k)$ with $k=2, \cdots, n{-}2$. $E_{n,k}$ counts the number of permutations of $\{1,2,\cdots, n\}$ with $k$ ascents, and it is obvious $\sum_{k=2}^{n{-}2} E_{n{-}3,k{-}2}=(n{-}3)!$. Some examples are for MHV and $\overline{\rm MHV}$ sectors, $E_{n{-}3, 0}=E_{n{-}3, n{-}4}=1$, $E_{3,1}=4$ ($n=6$), $E_{4,1}=E_{4,2}=11$ ($n=7$), and $E_{5,1}=E_{5,3}=26$, $E_{5,2}=66$ ($n=8$). 

We define the measure of $D=4$ twistor-string formula with these constraints:
\be\label{eq:dmu}
d\mu_{n,k}:=\frac{\prod_{i=1}^n t_i d t_i d\sigma_i}{{\rm vol.} SL(2)\times GL(1)} 
\delta^{2k} (C \cdot \tilde\lambda) 
\delta^{2(n{-}k)} (C^{\perp} \cdot \lambda),
\ee
where we have $2\times k {+} 2\times (n{-}k)=2n$ delta functions, and note that they contain the $4$ momentum-conserving delta functions, $\delta^4(\lambda \cdot \tilde\lambda)$, thus we have $2n{-}4$ delta functions for the $2n{-}4$ integration variables in $G(2,n)$. To see the derivation from \eqref{eq:polyrho} to \eqref{eq:cdl1}, note that the one only needs to integrate over $\rho$'s or $\tilde\rho$'s, respectively:
\begin{align}
&\delta^{2k} (C\cdot \tilde\lambda)=\int d^{2(n{-}k)} \tilde\rho \prod_i \delta^2 (\tilde\lambda_i-\tilde{t}_i \sum_{\tilde{m}}\tilde\rho_{\tilde{m}} \sigma_i^{\tilde{m}})\,,\nonumber\\
&\delta^{2(n{-}k)} (C^{\perp} \cdot \lambda)=\int d^{2k} \rho \prod_i \delta^2(\lambda_i - t_i \sum_m \rho_m \sigma_i^m)\,.
\end{align}

With this measure, we can write down twistor-string formulas for tree amplitudes in ${\cal N}=4$ SYM and those in ${\cal N}=8$ SUGRA, where $k=2,\cdots, n{-}2$ denotes the MHV degree; note that these super-amplitudes are degree-$(k {\cal N})$ polynomials of Grassmann variables $\eta^I_i$ with $I=1,\cdots, {\cal N}$: 
\begin{align}
& A^{{\cal N}=4}_{n,k}=\int d\mu_{n,k} \frac{\delta^{0|4k} (C \cdot \tilde\eta)}{(12) \cdots (n1)}\,,\nonumber\\
& A^{{\cal N}=8}_{n,k}=\int d\mu_{n,k} \delta^{0|8 k} (C \cdot \tilde\eta) R(\rho) \tilde{R} (\tilde\rho)\,,
\end{align}
where we have introduced the fermionic delta functions (in analog with those for $C \cdot \tilde\lambda$) which contains super-charge conserving delta functions $\delta^{0|2{\cal N}} (\sum_i \lambda_i \tilde\eta_i)$; for SYM case the integrand is the ``Parke-Taylor" factor, or cyclic measure of $G(2,n)$, with $(i j):=t_i t_j \sigma_{i,j}$, and for SUGRA the integrand is given by the resultants of the maps, $R(\rho)$ and $R(\tilde{\rho})$~\cite{Cachazo:2013zc}. 

Similarly such twistor-string formulas for SYM and SUGRA (with maximal or reduced supersymmetries) in $D=3$ have been obtained in~\cite{Cachazo:2013iaa} by considering a dimension reduction; more details about the scattering equations and formulas in $D=3$ can be found there. While for SYM we still have all sectors, it was observed in~\cite{Cachazo:2013iaa} that for SUGRA, only the middle sector $d=\tilde{d}$, or $k=d{+}1=n/2$ survives the dimension reduction (and only with even $n$); this is also the only sector for $D=3$ where solutions are all distinct (without multiplicities). Instead of Eulerian numbers, we have tangent number (or Euler zag number) $E_{n{-}3}$ of solutions (in the middle sector $k=n/2$) defined as $$\tan(x)=\sum_p \frac{E_{2p-1} x^{2p-1}}{(2p{-}1)!}=x+\frac{2 x^3}{3!}+ \frac{16 x^4}{5!}+\cdots,$$
thus there are {\it e.g.} $1,2,16$ solutions for $n=4,6,8$, respectively. 

Remarkably, similar formulas for amplitudes in ABJM and BLG, which also only exist in the middle sector $k=n/2$ for even $n$, naturally come about in this setting. Recall that 
ABJM theory is a supersymmetric extension of 3d Chern-Simons matter theory with supercharge $\mathcal{N}=6$. The R-symmetry is $SO\left(6\right)=SU\left(4\right)$ and the physical degrees of freedom are 4 complex scalars $X_{\mathrm{A}}$ and 4 complex fermions $\psi^{\mathrm{A}\alpha}$ as well as their complex conjugates $\bar{X}_{\mathrm{A}}$ and $\bar{\psi}_{\mathrm{A}\alpha}$. They transform in the fundamental or anti-fundamental of $SU\left(4\right)$ and $\mathrm{A}=1,2,3,4$ and $\alpha=1,2$ is the spinor index. These states can be arrange by in on-shell superspace by introduce three anti-commuting variables $\eta_{A}$ with $A=1,2,3$:
\begin{align}
	\Phi=X_{4}+\eta_{A}\psi^{A}-\frac{1}{2}\epsilon^{ABC}\eta_{A}\eta_{B}X_{C}-\eta_{1}\eta_{2}\eta_{3}\psi^{4},\\\nonumber
\bar{\Psi}=\bar{\psi}_{4}+\eta_{A}\bar{X}^{A}-\frac{1}{2}\epsilon^{ABC}\eta_{A}\eta_{B}\bar{\psi}_{C}-\eta_{1}\eta_{2}\eta_{3}\bar{X}^{4}.
\end{align}
We have split the fields as $X_{\mathrm{A}}\rightarrow\left(X_{4},X_{A}\right)$ and $\psi^{\mathrm{A}}\rightarrow\left(\psi^{4},\psi^{A}\right)$, and similarly for $\bar{X}^{\mathrm{A}}$ and $\bar{\psi}_{\mathrm{A}}$. There are two classes of amplitudes $\mathcal{A}_{n}\left(\bar{1}2\bar{3}\cdots n\right)$ and $\mathcal{A}_{n}\left(1\bar{2}3\cdots\bar{n}\right)$; 
under the little group transformation, $\lambda_{i}^{\alpha}\rightarrow -\lambda_{i}^{\alpha}$ and $\eta_{i}^{\mathrm{A}}\rightarrow -\eta_{i}^{\mathrm{A}}$, 
the amplitude with $\Phi_i$-supermultiplet is invariant while that with $\bar{\Psi_i}$-supermultiplet will pick up the minus sign. The super-amplitude has Grassmann degree $3 k=3 n/2$. What will be more relevant to our discussion later is to consider amplitudes with reduced SUSY in the positive branch. We only introduce two anti-commuting variables $\eta^\text{A}$ with $\text{A}=1,2$, and physical states can be rearranged by four supermultiplets $\Phi(i)$, $\bar{\Psi}(i)$ for $i=1,2$ as follow 
\begin{equation}
\begin{split}
    &\Phi\left(1\right):=\psi^{3}-\frac{1}{2}\epsilon^{ab}\eta_{a}X_{b}-\eta_{1}\eta_{2}\psi^{4},\qquad\Phi\left(2\right):=X_{4}+\eta_{a}\psi^{a}+\eta_{1}\eta_{2}X_{3}\\
    &\bar{\Psi}\left(1\right)=\bar{X}^{3}-\frac{1}{2}\epsilon^{ab}\eta_{a}\bar{\psi}_{b}-\eta_{1}\eta_{2}\bar{X}^{4},\qquad \bar{\Psi}\left(2\right)=\bar{\psi}_{4}+\eta_{a}\bar{X}^{a}-\frac{1}{2}\epsilon^{ab}\eta_{a}\eta_{b}\bar{\psi}_{3}
\end{split}
\end{equation}
$\Phi(1)$ and $\bar{\Psi}(1)$ can identify from $\Phi$ and $\bar{\Psi}$ by integrating out $\eta^3$, while $\Phi(2)$ and $\bar{\Psi}(2)$ can identify from $\Phi$ and $\bar{\Psi}$ by setting $\eta^3=0$. Now we only consider the sector, the odd sites with the  supermultiplet $\Phi(1)$ and even sites with  supermultiplet $\bar{\Psi}(2)$. This sector in term of Grassmannian integral will be equivalent to integrate out the $\eta^3$ on the odd sites 
$$A_{n}^{\rm 3d, reduced}=	\int\prod_{{\rm odd} i} d\eta_{i}^{3} A_n^{\rm 3d}. $$

Let us record the twistor-string-like formula for ABJM amplitude proposed by Huang and Lee in~\cite{Huang:2012vt}, as an integral over $G(2,n)$:
\begin{equation}
	A^{\rm 3d}_{n}=\int \frac{d^{2n}C}{\mathrm{vol}\,GL\left(2\right)}\frac{J\,\Delta\,\delta^{2k|3k}\left(C\cdot (\lambda|\eta) \right)}{\left(12\right)\left(23\right)\cdots\left(n1\right)}
\end{equation}
where we have $D=3$ spinors $\lambda_i^\alpha$ with $\alpha=1,2$ and Grassmann variables $\eta_{i}^{A}$ with $A=1,2,3$, and $\left(i\,j\right)=a_{i}b_{j}{-}a_{j}b_{i}$; we have $2d{+}1=n{-}1$ delta functions
$$\Delta=\prod_{j=1}^{2d{+}1}\delta\left(\sum_{i=1}^n (-)^{i{-}1} a_{i}^{2d{+}1{-}j}b_{i}^{j{-}1}\right)$$
(note $d=k{-}1$), the $k\times 2k$ $C$ matrix and the Jacobian
$$
C=\left(\begin{array}{cccc}
a_{1}^{d} & a_{2}^{d} & \cdots & a_{n}^{d}\\
a_{1}^{d{-}1}b_{1} & a_{2}^{d{-}1}b_{2} & \cdots & a_{n}^{d{-}1}b_{n}\\
\vdots & \vdots &  & \vdots\\
a_{1}b_{1}^{d{-}1} & a_{2}b_{2}^{d{-}1} & \cdots & a_{n}b_{n}^{d{-}1}\\
b_{1}^{d} & b_{2}^{d} & \cdots & b_{n}^{d}
\end{array}\right) \qquad 	J=\frac{\prod_{1\leq i<j\leq2d{+}1}\left(ij\right)}{\prod_{1\leq i<j\leq d{+}1}\left(2i{-}1,2j{-}1\right)}.
$$
Note that the $C$ matrix is also a Veronese map from $\{(a_i, b_i)|i=1,\cdots, n\} \in G(2,n)$ to $G(k, n=2k)$, and the $n{-}1$ delta functions in $\Delta$ encode their orthogonal conditions 
\begin{equation}    
C(\{a,b\}) \cdot \Omega \cdot C^{\rm T}(\{a,b\})=0\,,
\end{equation}
where $\Omega=(+, -, +,-,\cdots)$ is the metric such that we can have $C\in G_+(k,2k)$. As will be discussed in appendix A, ABJM tree amplitudes and leading singularities are encoded in positive orthogonal Grassmannians $OG_+(k,2k)$~\cite{Huang:2013owa,Huang:2014xza}. The $n$ delta functions imposing $C(\{a,b\})\cdot \lambda=0$ are essentially $D=3$ scattering equations, which implies momentum conservation (for $\alpha,\beta=1,2$ and symmetric)
$$
(\lambda \cdot \Omega \cdot \lambda)^{\alpha \beta}=\sum_{i=1}^n (-)^{i{-}1} \lambda^\alpha_i \lambda^\beta_i=0. 
$$ 

As will be discussed in section~\ref{sec:3d}, there is a crucial difference between $D=4$ and $D=3$: while the former has little group $U(1)$, the latter has only $Z_2$. In $D=4$ it is natural to have $t_i$ (and $\tilde t_i \propto 1/t_i$) which combines with $\{\sigma_i\} \in {\cal M}_{0,n}$ into the $D=4$ moduli space $G(2,n)$; in $D=3$ such $t_i$'s are fixed (up to a sign, or $\mathbb{Z}_2$ redundancy) by the $n{-}1$ orthogonality conditions of the Veronese $OG(k,n=2k)$. This allows us to rewrite twistor-string formula for ABJM as integrals over $\{\sigma_i\} \in {\cal M}_{0,n}$, which is the $D=3$ moduli space, as discussed in~\cite{Cachazo:2013iaa}. What we will find is a natural map and pushforward from ${\cal M}_{0,n}^+$ (as opposed to $G_+(2,n)$) to the $(n{-}3)$-dim momentum amplituhedron of ABJM.

\section{Twistor-string map in $D=4$ and the momentum amplituhedron of SYM}\label{4d}
In this section, after reviewing the momentum amplituhedron of SYM whose canonical forms give super-amplitudes, we propose a $D=4$ twistor-string map and conjecture that it provides a diffeomorphism from $G_+(2,n)$ to the interior of the momentum amplituhedron.  We also study how boundaries are mapped and how helicity amplitudes are obtained by pullback to subspaces. 
\subsection{The momentum amplituhedron of ${\cal N}=4$ SYM}
\paragraph{Definitions} The momentum amplituhedron  $\mathcal{M}(k,n)$ is the positive geometry associated with N$^{k-2}$MHV tree-level amplitude in $\mathcal{N}=4$ SYM~\cite{Damgaard:2019ztj}, which recast the original amplituhdedron~\cite{Arkani-Hamed:2013jha} in momentum-twistor space to spinor- helicity space. In the momentum amplituhedron, the kinematic data is bosonized by introducing $2(n-k)$ Grassmann-odd variables $\phi^{\alpha}_a$, $\alpha=1,\ldots n-k$ and $2k$ Grassmann-odd variables $\tilde{\phi}^{\dot{\alpha}}_{\dot{a}}$, $\dot{\alpha}=1,\ldots k$ and as

\begin{equation}
	\Lambda_{i}^{A}=\left(\begin{array}{c}
\lambda_{i}^{a}\\
\phi_{a}^{\alpha}\cdot\eta_{i}^{a}
\end{array}\right),\quad A=\left(a,\alpha\right)=1,\ldots,n-k+2,
\end{equation}

\begin{equation}
	\tilde{\Lambda}_{i}^{\dot{A}}	=\left(\begin{array}{c}
\tilde{\lambda}_{i}^{a}\\
\tilde{\phi}_{a}^{\alpha}\cdot\tilde{\eta}_{i}^{a}
\end{array}\right),\quad\dot{A}=\left(\dot{a},\dot{\alpha}\right)=1,\ldots,k+2.
\end{equation}
We introduce the matrices $\Lambda, \tilde{\Lambda}$ as follows, and ask $\tilde{\Lambda}$ ($\Lambda$) to be (twisted-)positive:
\begin{equation}
	\Lambda:=\left(\begin{array}{cccc}
\Lambda_{1}^{A} & \Lambda_{2}^{A} & \ldots & \Lambda_{n}^{A}\end{array}\right)\in M^+_{\rm twisted}\left(n-k+2,n\right),\quad\tilde{\Lambda}=\left(\begin{array}{cccc}
\tilde{\Lambda}_{1}^{\dot{A}} & \tilde{\Lambda}_{2}^{\dot{A}} & \ldots & \tilde{\Lambda}_{n}^{\dot{A}}\end{array}\right)\in M^+\left(k+2,n\right)\nonumber
\end{equation}
where $\Lambda$ is twisted-positive means that its orthogonal complement is positive $\Lambda^{\perp} \in M^+(k{-}2, n)$ (see~\cite{Damgaard:2019ztj}), and we refer to $(\Lambda,\tilde{\Lambda})$ as the kinematic data. 
Given such ``positive kinematic data", there are two ways to define the momentum amplituhedron.~\footnote{The first definition does not guarantee that the Mandelstam variables
are always positive for all points in the image of $G_+(k,n)$ through the map $\tilde{\Phi}_{\Lambda,\tilde{\Lambda}}$ while the second explicitly assume this. In order to agree with the second definition, we need to impose extra conditions on $\Lambda$ and $\tilde{\Lambda}$ to ensure these positivity conditions~\cite{Damgaard:2019ztj}.} The first is by introducing auxiliary positive Grassmannian matrix $C\in G_+(k,n)$ and we define momentum amplituhedron   ${\cal M}(k,n)$ as the image of the following map:
\begin{equation}\label{eq:def of momentum amplituhedron}
	\tilde{\Phi}_{\Lambda,\tilde{\Lambda}}:G_{+}(k,n)\to G(n{-}k,n{-}k{+}2)\times G(k,k{+}2):\quad Y_{a}^A=c_{\alpha,i}^\perp \Lambda_i^A\qquad\tilde{Y}_{\dot{a}}^{\dot{A}}=c_{\dot{\alpha},i} \Lambda_i^{\dot{A}}.
\end{equation}
After moduling out momentum conservation, it is easy to see that ${\cal M}(k, n)$ lives in a $2n{-}4$ dimension subspace of $G(n{-}k,n{-}k{+}2)\times G(k,k{+}2)$ satisfying~\footnote{Alternatively, the momentum conservation can totally express in term of $\left\langle Yij\right\rangle$ and $[\tilde{Y}ij]$ via 
$\frac{1}{\left\langle Y12\right\rangle ^{2}[\tilde{Y}12]^{2}}\sum_i\left\langle Yia\right\rangle [\tilde{Y}ib]$ with $a,b=1,2$.}
\begin{equation}\label{eq:momcon}
P^{a \dot{a}}=\sum_{i=1}^{n}\left(Y^{\perp} \cdot \Lambda\right)_{i}^{a}\left(\widetilde{Y}^{\perp} \cdot \widetilde{\Lambda}\right)_{i}^{\dot{a}}=0 .
\end{equation}

Just as the amplituhedron in momentum-twistor space~\cite{Arkani-Hamed:2017vfh}, the definition of the momentum amplituhedron implies particular sign patterns~\cite{Damgaard:2019ztj}. One can show that the brackets $\left\langle Yii{+}1\right\rangle$, $[\tilde{Y}ii+1]$  are positive and the sequence $\{\langle Y 12\rangle,\langle Y 13\rangle, \ldots,\langle Y 1 n\rangle\}$
has $k{-}2$ sign flips and the sequence $\{[\tilde{Y} 12],[\tilde{Y} 13], \ldots,[\tilde{Y} 1 n]\}$ has $k$ sign flips.

The second way for defining the momentum amplituhedron~\cite{Damgaard:2020eox} is directly in kinematic space by projecting the kinematic data through $Y,\tilde{Y}$
\begin{equation}\label{eq:kinematic momentum amplituhedron 1}
\lambda_{i}^{a}=\left(Y^{\perp}\right)_{A}^{a} \Lambda_{i}^{A} \quad \widetilde{\lambda}_{i}^{\dot{a}}=\left(\widetilde{Y}^{\perp}\right)_{\dot{A}}^{\dot{a}} \widetilde{\Lambda}_{i}^{\dot{A}}
\end{equation}
With GL$(n-k)$ and GL$(k)$ gauge redundancy of $Y$ and $\tilde{Y}$, we can gauge fixing them as  \begin{equation}\label{eq:gaugeY}
Y_{\alpha}^{A}=\left(\begin{array}{c}
{-}y_{\alpha}^{a} \\
1_{(n{-}k) \times(n{-}k)}
\end{array}\right), \quad \widetilde{Y}_{\dot{\alpha}}^{\dot{A}}=\left(\begin{array}{c}
{-}\widetilde{y}_{\dot{\alpha}}^{\dot{\alpha}} \\
1_{k \times k}
\end{array}\right)
\end{equation}
Then their orthogonal complements are
\begin{equation}\label{eq:kinematic momentum amplituhedron 2}
	Y^{\perp}_{(n{-}k)\times(n{-}k{+}2)}=\left(I_{2\times2}\:\tilde{y}_{2\times(n{-}k)}\right), \qquad \tilde{Y}_{2\times(k{+}2)}^{\perp}=\left(I_{2\times2}\:\tilde{y}_{2\times k}\right)
\end{equation}
Moreover, we can decompose the matrices $\Lambda,\tilde{\Lambda}$ according \begin{equation}\label{eq:kinematic momentum amplituhedron 3}
\Lambda_{i}^{A}=\left(\begin{array}{c}
\lambda_{i}^{a *} \\
\Delta_{i}^{\alpha}
\end{array}\right), \quad \widetilde{\Lambda}_{i}^{\dot{A}}=\left(\begin{array}{c}
\widetilde{\lambda}_{i}^{\dot{a} *} \\
\widetilde{\Delta}_{i}^{\alpha}
\end{array}\right)
\end{equation}
where $\lambda,\tilde{\lambda}^*$ are fixed $2$-planes in $n$ dimensions, and $\Delta,\tilde{\Delta}$ are fixed $(n-k)$-plane and $k$-plane and we assume that $\Lambda$ is a twisted positive matrix and $\tilde{\Lambda}$ is a positive matrix. 
After plugging~\eqref{eq:kinematic  momentum amplituhedron 2} and~\eqref{eq:kinematic  momentum amplituhedron 3} into~\eqref{eq:kinematic  momentum amplituhedron 1}, we arrive at  
\begin{equation}\label{eq:gaugel}
\mathcal{V}_{k,n}=\left\{(\lambda_{i}^{a}, \widetilde{\lambda}_{i}^{\dot{a}}): \lambda_{i}^{a}=\lambda_{i}^{* a}{+}y_{\alpha}^{a} \Delta_{i}^{\alpha}, \widetilde{\lambda}_{i}^{\dot{a}}=\widetilde{\lambda}_{i}^{* \dot{a}}{+}\widetilde{y}_{\dot{\alpha}}^{\dot{a}} \widetilde{\Delta}_{i}^{\dot{\alpha}}, \sum_{i=1}^{n}\lambda_{i}^{a} \widetilde{\lambda}_{i}^{\dot{a}}=0\right\}
\end{equation}
Note that $\mathcal{V}_{k,n}$ is a co-dimension-four subspace of an affine space of dimension $2n$. We also define a winding space $\mathcal{W}_{k,n}$
\begin{align}
\label{winding.definition}
\mathcal{W}_{k,n}\equiv &\{(\lambda_i^a,\widetilde\lambda_i^{\dot a}):\langle i\ i+1\rangle\geq 0,[i\ i+1]\geq 0, s_{i,i+1,\ldots,i+j} \geq 0\,, \nonumber \\&
\mbox{the sequence } \{\langle 12\rangle,\langle 13\rangle,\ldots,\langle 1n\rangle\} \mbox{ has } k-2 \mbox{ sign flips}\,, \nonumber \\
&\mbox{the sequence } \{[ 12],[ 13],\ldots,[ 1n]\} \mbox{ has } k \mbox{ sign flips}\} \,,
\end{align}
Then $\mathcal{M}(k,n)$ in 4d kinematic space is defined as the intersection:
\begin{equation}\label{eq:2nm4}
	\mathcal{M}(k,n)\equiv\mathcal{V}_{k,n}\cap\mathcal{W}_{k,n}
\end{equation}
There is a natural relation of the brackets in $Y,\tilde{Y}$ space and those in the kinematics space
$\left\langle Yij\right\rangle \rightarrow\left\langle ij\right\rangle$, $[\tilde{Y}ij]\rightarrow[ij]$
which we derive in appendix~\ref{sec:bracket} for completeness. Therefore, for any point in ${\cal M}(k,n)$ according to the first definition, the point in kinematic space must have correct sign flips and positivity in the second definition (and vice versa), and the two definitions can be easily translated.~\footnote{Only when certain conditions already impose on $\Lambda$ and $\tilde{\Lambda}$ guarantee that the Mandelstam variables
are always positive.} 

The boundaries of the momentum amplituhedron were studied in~\cite{Damgaard:2019ztj} and~\cite{Ferro:2020lgp}, which correspond to singularities of  SYM amplitudes. In particular,  the co-dimension one boundaries of the momentum amplituhedron are
\begin{equation}
\langle Y i i+1\rangle=0, \quad[\tilde{Y} i i+1]=0, \quad S_{a, a+1, \ldots, b}=0~(b-a>1)
\end{equation}
where the first two classes correspond to collinear limits and the last class correspond to multi-particle factorizations of the amplitude~\footnote{We have denoted Mandelstam variables in general $D$ dimensions as $s_{i,j}$; in $D=4$ we use $S_{i,j}:=\langle Y i j\rangle [\tilde{Y} i j]$ instead, which equals $s_{i,j}$ up to an overall constant (similarly for $D=3$), and planar Mandelstam variables are defined as $S_{a, a{+}1, \cdots, b}:=\sum_{a\leq i<j\leq b} S_{a,b}$.}.

However, although the collinear boundary can be proved easily, the factorization boundary is nontrivial except for $k=2, n-2$~\cite{Damgaard:2019ztj}. For $2<k<n{-}2$, it requires the external data $\Lambda$ and $\tilde{\Lambda}$ to satisfy some extra condition. An example to realize the factorization boundary is that $\Lambda$ and $\tilde{\Lambda}$ is on the momentum curve, i.e.
\begin{equation}\label{eq:possL}
\left(\Lambda^{\perp}\right)_{i}^{\bar{A}}=i^{\bar{A}-1}, \quad \tilde{\Lambda}_{i}^{\dot{A}}=i^{\dot{A}-1}
\end{equation}

\paragraph{Canonical forms} Having defined the space $\mathcal{M}(k,n)$, we want to write down its volume form i.e. the differential form with logarithmic singularities on the boundaries of $\mathcal{M}(k,n)$. The volume form will be related to the scattering amplitude of $\mathcal{N}=4$ SYM theory. The common way to obtain the volume form is by push-forward the BCFW cell to the momentum amplituhedron and wedge $\delta^{4}\left(P\right)d^{4}P$ to make it become top invariant form. Push forward can be either chose to $(Y,\tilde{Y})$ space or to the spinor helicity space. 

First, we discuss push forward canonical form of a set of BCFW cells with $(2n-4)$ degree of freedom to the $(Y,\tilde{Y})$ space. For example $n=4$, $k=2$
\begin{equation}
	C=\left(\begin{array}{cccc}
1 & x_{2} & 0 & -x_{3}\\
0 & x_{1} & 1 & x_{4}
\end{array}\right)
\end{equation}
The push-forward $(2n-4)$ dlog form of the Grassmannian $C$ through~\eqref{eq:def of momentum amplituhedron} is
\begin{equation}
	\boldsymbol{\Omega}_{n,k}=\bigwedge_{j=1}^{4}\mathrm{dlog}\,x_{j}=\mathrm{dlog}\frac{\left\langle Y12\right\rangle }{\left\langle Y13\right\rangle }\land\mathrm{dlog}\frac{\left\langle Y23\right\rangle }{\left\langle Y13\right\rangle }\land\mathrm{dlog}\frac{\left\langle Y34\right\rangle }{\left\langle Y13\right\rangle }\land\mathrm{dlog}\frac{\left\langle Y14\right\rangle }{\left\langle Y13\right\rangle }
\end{equation}
After multiplying  $\boldsymbol{\Omega}_{n,k}$ with $\delta^4 (P) d^4 P$, we can obtain the volume function $\Omega_{n,k}$. Since the volume form $\delta^4 (P) d^4 P \land\, \boldsymbol{\Omega}_{n,k}$ is ($2n$-dim) top form in $(Y,\tilde{Y})$ space, it can be written in terms of the top measure of $G(k,k{+}2)\times G(n{-}k,n{-}k{+}2)$, multiplied by a volume function $\Omega_{n,k}$; for the $n=4$ case we have $\Omega_{n,k}\left\langle Yd^{2}Y_{1}\right\rangle \left\langle Yd^{2}Y_{2}\right\rangle [\tilde{Y}d^{2}\tilde{Y}_{1}][\tilde{Y}d^{2}\tilde{Y}_{2}]$
with
\begin{equation}\label{eq:4ptMHV}
	\Omega_{n,k}=\frac{\delta^4(P)\left\langle 1234\right\rangle ^{2}\left[1234\right]^{2}}{\left\langle Y12\right\rangle \left\langle Y23\right\rangle [\tilde{Y}12][\tilde{Y}23]}
\end{equation}
Similar to the ordinary amplituhedron, we can extract the amplitude from the volume function $\Omega_{n,k}$. The procedure is to localize the $Y$ and $\tilde{Y}$ on the reference subspace 
 \begin{equation}
 Y^*=\left(\begin{matrix}
0_{2\times (n-k)}\\
\hline
1_{(n-k)\times (n-k)}
\end{matrix}\right),\qquad\qquad 
\tilde Y^*=\left(\begin{matrix}
0_{2\times k}\\
\hline
1_{k\times k}
\end{matrix}\right) ,
\end{equation}  
obtaining $n$-point N${}^{k{-}2}$MHV super-amplitude in non-chiral superspace (with ${\cal N}=(2,2)$):
 \begin{equation}
 \label{extract}
A^{\tt tree}_{n,k}=   \delta^4(p) \int d \phi^1_a \ldots  d\phi^{n-k}_a\int d\tilde\phi^1_{\dot a}\ldots  d\tilde\phi^k_{\dot a}\, \,\Omega_{n,k}(Y^*,\tilde Y^*,\Lambda,\tilde\Lambda) \,,
 \end{equation}
Using this procedure, we can obtain $n=4$ amplitude from~\eqref{eq:4ptMHV}:
\begin{equation}
	\frac{\delta^{4}\left(P\right)\delta^{4}(q)\delta^{4}(\tilde{q})}{\left\langle 12\right\rangle \left\langle 23\right\rangle \left[12\right]\left[23\right]}
\end{equation}

In general, we pushforward canonical forms of BCFW cells $\omega_{n,k}^{(\gamma)}$ to kinematic space~\cite{He:2018okq}:
\begin{equation}
	\bold{\Omega}_{n,k}^{(\gamma)}=\int\bold{\omega}_{n,k}^{(\gamma)}\prod_{\mu'}\delta^{2}\left(C_{\mu'}\left(x\right)\cdot\tilde{\lambda}\right)\prod_{\tilde{\mu}}\delta^{2}\left(C_{\tilde{\mu}}^{\perp}\left(x\right)\cdot\lambda\right)\bigwedge_{\mu'}\left(C_{\mu'}\left(x\right)\cdot d\tilde{\lambda}\right)^{2}\bigwedge_{\tilde{\mu}}\left(C_{\tilde{\mu}}^{\perp}\left(x\right)\cdot d\lambda\right)^{2}
\end{equation}
This pushforward formula is very similar to the usual Grassmannian integral formula. A slight difference is it replaces the Grassmann variables $\eta_i$, $\tilde{\eta_i}$ with $d\lambda_i$, $d\tilde{\lambda_i}$. One can extract the amplitude from the form by multiplying the form $\bold{\Omega}_{n,k}^{(\gamma)}$ with $\delta^4(P)d^4P$ and make the replacement $d\lambda_i$ and $d\tilde{\lambda_i}$ to $\eta_i$ and $\tilde{\eta_i}$
\begin{equation}
A_{n,k}^{\rm tree}=	\left.\delta^{4}(P)d^{4}P\land\bold{\Omega}_{n,k}^{(\gamma)}\right|_{d\lambda_{i},d\tilde{\lambda}_{i}\rightarrow\eta_{i},\tilde{\eta}_{i}}.
\end{equation}

\subsection{Twistor-string map in $D=4$ and the main conjecture}
Let us define the twistor-string map in $D=4$. First we have the {\it Veronese map} from the $2n{-4}$ dimensional moduli space $G_+(2,n)$ to $G_{+}(k,n)$~\cite{Arkani-Hamed:2009kmp}. Given a point $c_i^{\alpha}=(t_i,\sigma_i),i=1,2,\dots,n$ in $G_{+}(2,n)$, the Veronese map gives a point in $G_+(k,n)$ via (for $\alpha=1,\cdots, k$)
\begin{equation}\label{eq:CV}
C(\sigma,t)_{\alpha, i}=t_i \sigma^{\alpha-1}_i
\end{equation}
Using Vandermende determinant, it is easy to see that the orthogonal complement $C^{\perp}\in G_+(n{-}k, n)$ can be written as
\begin{equation}\label{eq:CVP}
C^{\perp}(\sigma,t)_{\tilde{\alpha}, i}=\tilde{t}_i \sigma_i^{\tilde{\alpha}-1}
\end{equation}
where $\tilde{\alpha}=1,\cdots, n{-}k$ and  $\tilde{t}_i:=\frac{1}{t_i\prod_{j\neq i}(\sigma_i-\sigma_j)}$.

Given any kinematic point $\Lambda$ and $\tilde{\Lambda}$, where $\Lambda$ is a twisted positive $(n{-}k{+}2)\times n$ matrix and $\tilde{\Lambda}$ is a positive $(k{+}2)\times n$ matrix as defined above, we define the {\it twistor-string map} from $G_{+}(2,n)$ to $G(k, k{+}2)\times G(n{-}k, n{-}k{+}2)$ as
\begin{equation}\label{eq:YYts}
\Phi_{\Lambda, \tilde{\Lambda}}: (Y, \tilde{Y})=(C^{\perp}(\sigma, t) \cdot \Lambda, C(\sigma,t)\cdot \tilde\Lambda)
\end{equation} 
where the $C(\sigma,t)$ and $C^\perp(\sigma,t)$ are the Veronese matrix and its orthogonal complement given in~\eqref{eq:CV} and~\eqref{eq:CVP}. 
As we have reviewed, $Y \cdot Y^{\perp}=0$ gives  4 constraints on $(Y,\tilde{Y})$ thus the $(Y,\tilde{Y})$-space is of dimension $2n{-}4$ as expected. Equivalently, we can formulate the twistor-string map directly in the aforementioned $(2n{-}4)$-dim subspace of the spinor-helicity space. With gauge fixing in \eqref{eq:gaugeY} and \eqref{eq:kinematic momentum amplituhedron 2} 
we can rewrite the map in terms of $D=4$ scattering equations
\begin{equation}\label{eq:cdl}
C(\sigma, t)\cdot \tilde\lambda(y)=0,  C^\perp (\sigma, t) \cdot \lambda(\tilde y)=0
\end{equation}
which is a map from $G_{+}(2,n)$ to the $(2n{-}4)${-}dim $(y, \tilde{y})$ space (with $(\lambda^*, \tilde\lambda^*, \Delta, \tilde{\Delta})$ fixed).

As already pointed out in~\cite{Damgaard:2019ztj}, given any positive matrix in $G_{+}(k,n)$, $\tilde{\Phi}_{\Lambda,\tilde{\Lambda}}$ map gives the correct positive and the sign flip conditions. Therefore, given any point in $G_+(2,n)$, its image under our twistor-string must live inside the momentum amplituhedron. 

\paragraph{The main conjecture} The main conjecture of this section is that the twistor-string map~\eqref{eq:YYts} provides a diffeomorphism from $G_{+}(2,n)$ to the interior of momentum amplituhedron. In particular, the map provides a bijection between any point $(Y, \tilde{Y}) \in \mathcal{M}(k,n)$ and a point $(t_i,\sigma_i)\in G_{+}(2,n)$, which also implies that the image of $G_+(2,n)$ is exactly the momentum amplituhedron. As we discuss in the next subsection, a consequence of this conjecture is that the pushforward of $\bold{\Omega}(G_+(k,n))$ gives $\bold{\Omega}(\mathcal{M}(k,n))$, which have been proved in~\cite{Arkani-Hamed:2017tmz} and provides support and motivation for our conjecture. However, in the rest of the subsection, we will provide strong evidence for the main conjecture directly in terms of the positive geometries (not only their canonical forms).
 
What we need essentially is to show that the map is onto and one-one, {\it i.e.} given any point in $\mathcal{M}(k,n)$, out of the $E_{n{-}3,k{-}2}$ (generally complex) solutions of $D=4$ scattering equations, there is always a unique solution (pre-image of the $\Phi_{\Lambda,\tilde{\Lambda}}$ map) that is in $G_{+}(2,n)$. This is a very non-trivial statement about $D=4$ scattering equations and the momentum amplituhedron, for which we now provide some evidence. For the MHV and $\overline{\mathrm{MHV}}$ cases, where $E_{n{-}3,0}=E_{n{-}3,n{-}4}=1$ {\it i.e.} we have one solution. For $k=2$, the solution $t_i (1, \sigma_i)$ is (up to a GL$(2)$ transformation) proportional to $\lambda \in G_+(2,n)$; similarly for $k=n{-}2$, the solution is proportional to $\tilde\lambda \in G_+(2,n)$. Note that in either case, we need to require planar Mandelstam variables to be positive. These are the trivial cases where we have a diffeomorphism from $G_+(2,n)$ to $G_+(2,n)$.

For general $k$, we will provide highly non-trivial, strong numerical evidence that there is always a unique pre-image for any point in ${\cal M}(k,n)$, {\it i.e.} the twistor-string map is a one-to-one map.
To be more specific, we first generate data $(\lambda,\tilde{\lambda})$ with correct positivity and sign-flip condition: by starting with (twisted-)positive matrices  $\Lambda,\tilde{\Lambda}$, we compute points $(Y,\tilde{Y})$ in the momentum amplituhedron $\mathcal{M}(k,n)$ as the image via some points in (BCFW cells of) $G_{+}(k,n)$; then we derive $\lambda,\tilde{\lambda}$ from \eqref{eq:kinematic momentum amplituhedron 1} which by definition lie inside $\mathcal{M}(k,n)$. Note that although such external data points satisfy positivity and sign flip conditions, they do not need to give positive planar Mandelstam variables in general.  

With such external data at hand, we then solve for $D=4$ scattering equations. We have checked thoroughly up to $n=8$ with all $k$ sectors that there is always a unique pre-image in $G_+(2,n)$ for each point in ${\cal M}(k,n)$. There is a technical point we mention: computationally it is more convenient to first solve CHY scattering equations {\it e.g.} with gauge fixing $\sigma_1=0,\sigma_2=1,\sigma_n=\infty$ to get $(n{-}3)!$ solutions, then select the $E_{n-3,k-2}$ solutions in the correct $k$-sector and substitute them into \eqref{eq:cdl} to solve for $t_i$'s.  
Specifically, we have checked using the $3$ BCFW cells of $n=6,k=3$ (with $3\times 100$ points), $6$ cells of $n=7,k=3$ (with $6\times 10$ points), $10$ cells of $n=8,k=3$ (with $10\times 5$ points),and $20$ cells of $n=8,k=4$ (with $20\times 3$ points) (other non-trivial $k$ sectors can be obtained by parity $k \leftrightarrow n{-}k$). Regardless of whether the planar Mandelstam variables are positive or not, there is always a unique positive solution, out of $E_{n-3,k-2}$ solutions for each point.

We have not required planar Mandelstam variables to be positive. However, if we restrict ourselves to the momentum amplituhedron region where all of them are also positive, then from~\cite{Arkani-Hamed:2017mur} we have a unique solution (out of all $(n{-}3)!$ solutions) with $\sigma_1<\sigma_2\dots<\sigma_n$. For such kinematics, what is remarkable about our test is that for $D=4$ data with given sign flips, this solution is exactly in the corresponding $k$ sector, and all $t_i$ turn out to be positive! . On the other hand, we have also tested that for generic external data which do not obey positivity ($\langle i i{+}1\rangle$ or $\left[i i{+}1\right]$ not always positive) or sign-flip conditions: in these situations we do not have a unique positive solution. All these have strongly supported our conjecture that the map is indeed one-to-one. 

Last but not least, we remark that our conjecture is equivalent to the conjecture that the Jacobian of the map is non-vanishing as discussed in~\cite{Arkani-Hamed:2017tmz}. Although we do not have a proof, we can briefly recall the discussion of~\cite{Arkani-Hamed:2017tmz} and provide some numeric evidence for this claim. For any sector $k$, we start with $n=k{+}2$ where the conjecture holds as mentioned above. We proceed by induction: suppose for $n=m-1$, there is exactly one positive solution of \eqref{eq:cdl}; for $n=m$, we write matrix $C$ as a set of column vectors $C_i$ ($i=1,2\dots,n$) and
define two functions $\Delta_1,\Delta_2$ as follows
\begin{equation}
    \begin{split}
        \Delta_1(\gamma,C)&:=\gamma  C^\perp_1\cdot\lambda_1+\sum_{i=2}^m C^\perp\cdot\lambda_i\\
        \Delta_2(\gamma,C)&:=\gamma C_1\cdot\tilde{\lambda}_1+\sum_{i=2}^m C_i\cdot\tilde{\lambda}_i
    \end{split}
\end{equation}
and $\Delta(\gamma,C):=(\Delta_1(\gamma,C),\Delta_2(\gamma,C))$. When $\gamma=0$ the equation $\Delta(0,C)=0$ has unique positive solution by induction hypothesis, and what we want to show is that this remains true at $\gamma=1$. Note that points on the boundary of $G_+(2,n)$ do not satisfy the equations with generic kinematics, since they correspond to boundary (see below). Therefore, it suffices to show that the unique positive does not bifurcate as $\gamma$ evolves from $0$ to $1$; this is equivalent to the prove that the following (toric) Jacobian is non-vanishing for $0\leq \gamma \leq 1$. We denote $X_a=(t_i,\sigma_i)$ for $a =1,2,\dots,2n-4$ after fixing GL(2), and denote the $2n{-}4$ equations as $\Delta'_b$ where  $4$ equations are deleted due to momentum conservation:
\begin{equation}
    J(\gamma, X_a)=\mathrm{det}\left(X_a\frac{\partial \Delta'_b}{\partial X_a}\right)\neq 0, \quad 0\leq \gamma \leq 1
\end{equation}
We have numerically checked by choosing random points of $0\leq \gamma\leq 1$ and find that $J$ has definite sign for a given kinematic points. 

\paragraph{Boundaries} Another important question we consider is the map between boundaries of compactified $G_{+}(2,n)$ and those of $\mathcal {M}(k,n)$ under the twistor-string map. The fact that the co-dimension one boundaries can be nicely mapped to each other provides further support for our conjecture. These boundaries correspond to either multi-particle poles $X_{i,j}=s_{i,i{+}1, \cdots, j{-}1}= 0$ ($|j-i|>2$), or collinear poles, $\langle i i{+}1\rangle=0$ or $[ i i{+}1]=0$ of SYM amplitudes. The former correspond to co-dimension one boundaries of compactified ${\cal M}_{0,n}^+$, while the latter in addition depends on the behavior of $t$ (and $\tilde t$ variables). 

First recall that we can introduce $\frac{n(n-3)}{2}$ $u$ variables to compactify ${\cal M}^+_{0,n}=G_+(2,n)/ T$:
\begin{equation}
u_{i, j}=\frac{\left(\sigma_{i}-\sigma_{j{-}1}\right)\left(\sigma_{i{-}1}-\sigma_{j}\right)}{\left(\sigma_{i}-\sigma_{j}\right)\left(\sigma_{i{-}1}-\sigma_{j{-}1}\right)}=\frac{(i j{-}1)(i{-}1 j)}{(i j)(i{-}1 j{-}1)}\qquad 1 \leq i<j{-}1<n
\end{equation}
These variables satisfy the so-called $u$ equations, one for each variable:
\begin{equation}
1-u_{i, j}=\prod_{i<k<j<l} u_{k, l}.
\end{equation}
A non-trivial fact is that the solution space is $n{-}3$ dimensional; by requiring $0\leq u_{i,j}\leq 1$, they cut out the worldsheet associahedron with $n(n{-}3)/2$ co-dimension one boundary or (curvy) facets, and each of them is reached by sending a $u_{i j} \to 0$ (automatically forcing all incompatible $u_{k,l} \to 1$). As we have mentioned above (see discussions in~\cite{Arkani-Hamed:2017mur, Arkani-Hamed:2019mrd}), under the CHY scattering-equation map, each boundary $u_{i,j} \to 0$ (with $j>i{+}2$) is mapped to a facet of ABHY associahedron $X_{i,j}=s_{i,i{+}1, \cdots, j{-}1} \to 0$. All we need here is to ``uplift" the map to $D=4$: with our compactification the former becomes a boundary of $G_+(2,n)$, and via the $D=4$ map (which concerns the same positive solution with non-negative $t$'s) we obtain the latter as a co-dimension one boundary of ${\cal M}(k,n)$: we have $S_{i,i{+}, \cdots, j{-}1}\to 0$ but no other planar variables (including $\langle Y a a{+}1\rangle$, $[\tilde{Y} a a{+}1]$) vanish! More explicitly, we can express the Veronese map directly in terms of $u$'s and $t$'s, and check that by sending any $u_{i,j}\to 0$ we indeed have a co-dimension one boundary with only $S_{i,i{+}, \cdots, j{-}1}\to 0$ if we scale some $t$ variables appropriately. 

On the other hand, it becomes more subtle for $j=i{+}2$: as $u_{i,i{+}2}\to 0$ (or $\sigma_i$ and $\sigma_{i{+}1}$ pinches), we have $S_{i,i{+}1}\to 0$ but that only happens with either of the two collinear limits, $\langle Y i i{+}1\rangle\to 0$ or $[\tilde{Y} i i{+}1]\to 0$. In other words, with $u_{i, i{+}2} \to 0$, we have two possible co-dimension one boundaries of ${\cal M}(k,n)$, $\langle Y i i{+}1\rangle = 0$ or $[\tilde{Y} i i{+}1]= 0$ depending on how $t$ variables behave. From \eqref{eq:polyrho} we see that (hereafter we use brackets without $Y$ and $\tilde{Y}$)
\begin{equation}
\begin{split}
&\langle a b\rangle=(a b) P_{k{-}2}\left(\sigma_{a}, \sigma_{b}\right)\\
&\left[ a b\right]=\widetilde{(a b)} \tilde{P}_{n{-}k{-}2}\left(\sigma_{a}, \sigma_{b}\right),
\end{split}
\end{equation}
where $(ab)=t_a t_b\sigma_{ab},\widetilde{(ab)}=\tilde{t}_a\tilde{t}_b\sigma_{ab}$ and $P_{k{-}2}$ ($\tilde{P}_{n{-}k{-}2})$ is a homogeneous polynomial of $\sigma$'s with degree $k{-}2$ ($n{-}k{-}2)$. Therefore, if we send $\sigma_i \to \sigma_{i{+}1}$ while keeping their $t$'s finite, we reach at the co-dimension one boundary $\langle ii+1\rangle=0$; if instead we keep the $\tilde{t}$'s finite, we have the co-dimension one boundary $\left[ ii+1\right]=0$. 

However, for the two special cases of $k=2$ (MHV) and $k=n{-}2$ ($\overline{\rm MHV}$), we only have one type of collinear boundaries as co-dimension one boundaries. For MHV ($\overline{\rm MHV}$) case, $\left[i i{+}1\right]=0$ ($\langle i i{+}1\rangle=0$) are no longer co-dimension one boundaries of the momentum amplituhedron, and neither are multi-particle $S_{i, \cdots, j{-}1}=0$. This fact can be seen directly from the equation above: {\it e.g.} for $k=2$, the polynomial $P_{k{-}2}$ is a constant and the $C$-matrix $\{t_i (1,\sigma_i)| i=1,\cdots n \}$ is related to the $\{\lambda_i\}$ matrix by a GL(2) transformation. Therefore, as $\sigma_i, \sigma_{i{+}1}$ pinch, we immediately see that $\lambda_i, \lambda_{i{+}1}$ become proportional to each other thus $\langle i i{+}1 \rangle \to 0$, but we cannot reach those boundaries $[i i{+}1]\to 0$ at all. Moreover, as we send $u_{i, j}\to 0$ (with $j{-}i>2$) which means $\sigma_i, \cdots,  \sigma_{j-1}$ all pinch, we see immediately that all the corresponding $\lambda$'s become proportional, {\it i.e.} $\langle p p{+}1\rangle \to 0$ for $p=i, \cdots j-2$, thus the factorization channel $S_{i,i{+}1, \cdots, j{-}1}\to 0$ is a high co-dimension boundary of $\mathcal{M}(2,n)$. The same is true for ${\cal M}(n{-}2, n)$ with $\lambda$ and $\tilde{\lambda}$ swapped.

\subsection{From canonical forms to super- and helicity-amplitudes}
If our main conjecture holds, according to~\cite{Arkani-Hamed:2017tmz} there must be a pushforward formula from $\bold{\Omega}(G_{+}(2,n))$ to $\bold{\Omega}({\cal M}_{k,n}(Y, \tilde{Y}))$ by summing over all solutions:

\begin{equation}
\bold{\Omega}({\cal M}_{k,n}(Y, \tilde{Y}))=\sum_{\rm sol.}^{E_{n-3,k-2}} \frac{d^{2n{-}4} C}{(12)\cdots (n1)}
\end{equation}
where $(i j):=t_i t_j (\sigma_j{-}\sigma_i)$ and the measure of $G(2,n)$ is given by $d^{2n{-}4}C:=d^{2n}C/{\rm vol~GL}(2)$ as in~\eqref{eq:dmu}; together with the cyclic minors, it gives the canonical form of $G_+(2,n)$, which can also be written as $\bold{\Omega}({\cal M}_{0,n})$ times the $n{-}1$ form for $t$-part:
$$
\frac{d^{2n}C}{{\rm vol~GL}(2) (12)\cdots (n1)}=\frac{d^n \sigma}{{\rm vol~SL}(2)\sigma_{12}\cdots \sigma_{n1}}~ \frac 1 {{\rm vol~GL}(1)}~\prod_{i=1}^n \frac{d t_i}{t_i}. 
$$
Note that the $(Y, \tilde{Y})$ space has dimension $2n$, and the volume function $\Omega_{n,k}$, from which superamplitude can be extracted, is obtained as
\begin{equation}
\bold{\Omega}_{n,k} \wedge d^4 P =\Omega_{n,k} \prod_\alpha \langle Y_1 \cdots Y_{n{-}k} d^2 Y_\alpha \rangle \prod_{\dot{\alpha}} [ \tilde{Y}_1 \cdots \tilde{Y}_{n{-}k} d^2 \tilde{Y}_{\dot{\alpha}}]
\end{equation}
where we have used the fact that the momentum amplituhedron lives in a $(2n{-}4)${-}dim subspace with $\delta^4(P)$ implicitly on both sides. Note that the pushforward formula is an alternative way for computing the canonical forms and volume functions, which one usually computes by triangulating the positive geometries using {\it e.g.} BCFW cells. We can also write the pushforward without $(Y, \tilde Y)$ explicitly, but directly in the $(2n{-}4)$-dim subspace of the kinematic space (see \eqref{eq:2nm4}) as a form of $y, \tilde{y}$ variables:
\begin{equation}
\bold{\Omega}_{n,k}(y, \tilde y)=\sum_{\rm sol.} \frac{d^{2n{-}4}C}{(12)\cdots (n1)}
\end{equation} 
where the sum is over all solutions of $C\cdot \tilde\lambda=C^\perp \cdot \lambda=0$, and the only difference with that in $(Y, \tilde{Y})$ space is just the choice of variables. Exactly by pulling out the top{-}dim measure $d^{2k} y  d^{2(n{-}k)}\tilde{y}/(d^4 P)$, one arrives at the same volume function $\Omega_{n,k}$ written as
\begin{equation}
\Omega_{n,k}(y, \tilde{y})=\sum_{\rm sol.} \frac{1}{\det'\Phi~(12) \cdots (n1)}
\end{equation}
where $\Phi_{a,b}= \partial y_a/\partial x_b$ denotes the $2n\times 2n$ derivative matrix with $y_{1\leq a\leq 2n}$ denoting collectively $(y, \tilde{y})$ and $x_{1\leq b\leq 2n}$ denoting collectively $\{\sigma_i, t_i\}$; the notation ${\rm det}'\Phi$ means that we need to delete 4 rows (corresponding to modding out $d^4 P$) and 4 columns (for vol.~GL$(2)$) with usual compensating factors. 

Note that these pushforward formulas for canonical forms and volume functions are equivalent to RSV-Witten formulas for super-amplitudes, once we identify $\eta, \tilde{\eta}$ as $d\lambda, d\tilde{\lambda}$ as usual. Therefore the validity of these formulas does not depend on our conjecture, and they actually provide a strong support for the validity of the latter. 

Moreover, we can go further and ask how to extract helicity amplitudes directly, which is similar to extracting $m(\alpha|\beta)$ from the canonical form of associahedron by pullbacks. It turns out that we have a class of particularly simple choices of subspaces, where the pullback of the volume function gives helicity amplitudes. We focus on gluon amplitudes for simplicity ( other helicity amplitudes can be extracted similarly): for any $N \cup P=[n]$ with $k$ and $n{-}k$ labels for negative- and positive-helicity gluons respectively, we choose 
$\lambda^*_i=\lambda_i \delta_{i, N}$, $\tilde\lambda^*_i=\tilde\lambda_i \delta_{i,P}$ with $\delta_{i,N}=1$ for $i\in N$ and $0$ for $i\in P$ (similarly for $\delta_{i,P}$). We also choose $\Delta=I_{k\times k}$, $\tilde\Delta=I_{(n{-}k) \times (n{-}k)}$ such that $y_i=\lambda_i \delta_{i,P}$ and $\tilde{y}_i=\tilde\lambda_i \delta_{i,N}$. In other words, we choose the subspace where $\lambda_i$ with $i \in N$ and $\tilde{\lambda_i}$ with $i \in P$ are constants and the complement ones are variables spanning it. By using the $GL(k)$ transformation that brought the $k\times k$ submatrix of Veronese $C$ matrix with columns $i\in N$ to identity~\cite{Geyer:2014fka,He:2016vfi}, the $D=4$ scattering equations~\eqref{eq:cdl1} nicely become
$\tilde\lambda_{i \in N}=\sum_{j\in P} \frac{\tilde\lambda_j}{(i j)}$, $\lambda_{j \in P}=\sum_{i \in N} \frac{\lambda_i}{(j i)}$. Note that this choice of subspace does not give a positive geometry in general, but here we are only interested in pullback of the volume function; it is straightforward to see that the pullback gives exactly gluon helicity amplitude:
\begin{equation}
\Omega_{n,k}(N,P):=\sum_{\rm sol.} \frac{1}{{\rm det'} \Phi(N,P)~(12) \cdots (n1)}=A_{n,k}^{\rm gluon}(N,P)
\end{equation}
where the Jacobian is given by the (reduced) determinant of derivative matrix $\Phi(N,P)_{a,b}:=\partial y_a /\partial x_b$ with $y$ denoting collectively $\{\tilde\lambda_{i \in N}, \lambda_{j\in P}\}$. We recognize this formula as nothing but the $D=4$ ambitwistor-string formula for gluon amplitudes~\cite{Geyer:2014fka}:
\begin{equation}
A_{n,k}^{\rm gluon}(N,P)=\int \frac{d^{2n} \sigma}{{\rm vol.~GL}(2)(12) \cdots (n1)}\prod_{i \in N}  \delta^{2}(\tilde\lambda_i{-}\sum_{j\in P} \frac{\tilde\lambda_j}{(i j)}) \prod_{j \in P}\delta^{2}(\lambda_j{-}\sum_{i \in N} \frac{\lambda_i }{(j i)}) 
\end{equation}

\section{Twistor-string map in $D=3$ and the momentum amplituhedron of ABJM}\label{sec:3d}
In this section, we move to $D=3$ and find a natural twistor-string map from its moduli space, ${\cal M}_{0,n}^+$, to a new positive geometry which we conjecture to be the momentum amplituhedron for ABJM theory. Similar to the $D=4$ case, we conjecture that the $D=3$ twistor-strong map provides a diffeomorphism from ${\cal M}_{0,n}^+$ to the interior of ABJM momentum amplituhedron and the canonical form of the latter gives ABJM amplitudes with supersymmetries reduced.

\subsection{The ABJM momentum amplituhedron and reduced SUSY amplitudes}

Let us first propose a definition of ABJM momentum amplituhedron, which is very similar to that of SYM in $D=4$. In complete analogy we introduce bosonized kinematic variables with $n=2k$ Grassmannian-odd variables $\phi^{\alpha}_a$, $a=1,2$, $\alpha=1,\ldots k$:
\begin{equation}
	\Lambda_{i}^{A}=\left(\begin{array}{c}
\lambda_{i}^{a}\\
\phi_{a}^{\alpha}\cdot\eta_{i}^{a}
\end{array}\right),\qquad A=\left(a,\alpha\right)=1,\dots,k+2
\end{equation}
which is required to be a positive matrix $\Lambda \in M^+(2k, k{+}2)$~\footnote{In general, positive $\Lambda$ cannot guarantee that planar Mandelstam variables have definite signs inside the $\mathcal{M}_{k,2k}^{\text{3d}}$; we do not know such sufficient and necessary conditions. For our purposes, we can restrict $\Lambda$ on the moment curve where the sign of planar variables is definite.}. Then we define ABJM momentum amplituhedron $\mathcal{M}^{3d}(k,2k)$ as the image of the positive orthogonal Grassmannian $OG_{+}(k,2k)$ through a map
\begin{equation}
\label{Phi}
\tilde{\Phi}_{\Lambda}:OG_{+}(k,2k)\to G(k,k{+}2)\,,
\end{equation}
where for each element of positive orthogonal Grassmannian $C=\{c_{\dot\alpha i}\}\in OG_{+}(k,2k)$, we associate an element $Y\in G(k,k{+}2)$ via
\begin{align} \label{eq:ABJMhedron OG}
Y^A_\alpha=c_{\alpha i}\Lambda_i^A\,,
\end{align}

Similar to $D=4$ momentum amplituhedron map, this $D=3$ map $\tilde{\Phi}_\Lambda$ does not map $OG_{+}(k,2k)$ to the top dimension of $Gr(k,k{+}2)$; rather the image lives in the following co{-}dimension three surface in $G(k,k{+}2)$~\footnote{Alternatively, we can write these momentum-conservation conditions in $Y$ brackets as $\frac{1}{\left\langle Y12\right\rangle ^{3}}\sum_{i}(-1)^{i}\left\langle Yia\right\rangle \left\langle Yib\right\rangle=0$ for $a,b=1,2$.}:
 \begin{equation}\label{eq:mom con}
P^{a b}=\sum_{i,j=1}^n\left(Y^\perp \cdot \Lambda\right)^a_i \eta_{ij} \left( Y^\perp \cdot \Lambda \right)^{b}_j=0\,.
\end{equation} 

We can also define the ABJM momentum amplituhedron in the kinematic space by identifying $\lambda$ with $\Lambda$ projected through $Y$ by
\begin{equation}
\lambda_{i}^{a}\rightarrow Y^{\perp}\cdot \Lambda_{i}^{A}
\end{equation}
Accordingly, we gauge fixing $Y$ and decompose the matrix $\Lambda$ as 
\begin{equation}
Y_{\alpha}^{A}=\left(\begin{array}{c}
{-}y_{\alpha}^{\alpha} \\
1_{k \times k}\end{array}\right)
, \quad \Lambda_{i}^{A}=\left(\begin{array}{c}
\lambda_{i}^{a *} \\
\Delta_{i}^{\alpha}
\end{array}\right)
\end{equation}
where $\lambda^*$ is fixed $2$-planes in $n$ dimensions, and $\Delta$ is fixed $k$-plane and we assume that $\Lambda$ is a positive matrix. The $(n{-}3)$-dim subspace where the ABJM momentum amplituhedron lives in is
\begin{equation}\label{eq:fix3d}
\mathcal{V}_{k,2k}=\left\{\lambda_{i}^{a}: \lambda_{i}^{a}=\lambda_{i}^{* a}{+}y_{\alpha}^{a} \Delta_{i}^{\alpha}, \sum_{i=1}^{2k}(-1)^{i-1}\lambda_{i}^{a} \lambda_{i}^{\dot{a}}=0\right\}
\end{equation}
And of course  $\lambda_i$ needs to restrict in the positive region with correct winding condition~\footnote{ Here, we define planar Mandelstam variables in 3d as $s_{i,i+1,\cdots,j}=\sum_{i\leq p<q\leq j} (-1)^{p+q+1}\langle p q\rangle^2$.}
\begin{align}
\label{winding.definition3}
\mathcal{W}_{k,2k}\equiv &\left\{\lambda_i^a:\langle i\ i+1\rangle > 0, s_{i,i+1,\ldots,i+j} > 0\,,\right. \nonumber \\&
\left.\mbox{the sequence } \{\langle 12\rangle,\langle 13\rangle,\ldots,\langle 1n\rangle\} \mbox{ has } k \mbox{ sign flips}\right \}
\end{align}
Then the kinematic ABJM momentum amplituhedron is the intersection of the two spaces
\begin{equation}
	\mathcal{M}^{\text{3d}}(k,2k)\equiv\mathcal{V}_{k,2k}\cap\mathcal{W}_{k,2k}
\end{equation}

Note that we have defined a new positive geometry, and we will use the definition of~\eqref{eq:ABJMhedron OG} for computing the volume form in $n=4,6$ in $Y$ space, which turns out to give ABJM tree amplitudes with reduced SUSY. Our approach is pushing forward the canonical volume form  associated with BCFW cells of OG and uplifting it to $n$-form in $Y$ space. The volume function extracted by moving the measure $\prod_i\langle Y d^2 Y_i\rangle$ from the $n$-form is related to reduced SUSY amplitude in ABJM theory. Here we will focus on  the  case $n=4,6$, there are no triangulation of the cells. These cases are simpler. 

\paragraph{$n=4$ case} In this case, there is no extra condition imposing on the cells. We can directly push-forward the top cell. Here, we choose the cells in the cyclic gauge
\begin{equation}
	\left(\begin{array}{cccc}
c_{21} & 1 & c_{23} & 0\\
c_{41} & 0 & c_{43} & 1
\end{array}\right)
\end{equation}
 We can obtain these matrix element $c_{ij}$ from~\eqref{eq:ABJMhedron OG}. Due to the special kinematic of four point in 3d, there are many ways to express $c_{ij}$. One particular choice is
\begin{equation}
	c_{21}=\frac{\left\langle Y23\right\rangle }{\left\langle Y31\right\rangle },\quad c_{23}=\frac{\left\langle Y12\right\rangle }{\left\langle Y31\right\rangle },\quad c_{41}=\frac{\left\langle Y34\right\rangle }{\left\langle Y13\right\rangle },\quad c_{43}=\frac{\left\langle Y41\right\rangle }{\left\langle Y13\right\rangle }
\end{equation}
 The push-forward of the Grassmannian top form through~\eqref{eq:ABJMhedron OG} is therefore
 \begin{equation}\label{eq:canonical form ABJM}
 	\mathrm{dlog}\frac{\left\langle Y12\right\rangle }{\left\langle Y23\right\rangle }
 \end{equation}
 To uplift the canonical form~\eqref{eq:canonical form ABJM} to the top form, it requires wedge the form with $\delta^3(P)d^3P$. However, the image of the $Y$ solve the momentum conservation condition \eqref{eq:mom con}.
 
To write down $\delta^3(P)d^3P$, we can introduce three regulators $l_1,l_2,l_3$ to slightly break the momentum conservation in this way
\begin{equation}
	\left(\begin{array}{cccc}
l_{1}c_{21} & 1 & l_{3}c_{23} & 0\\
l_{2}c_{41} & 0 & c_{43} & 1
\end{array}\right)
\end{equation} 
Using~\eqref{eq:mom con}, $\delta^3(P)d^3P$ can be expressed as
\begin{equation}
	\delta^3(P) J_p dl_1 dl_2 dl_3
\end{equation}
Then the volume form is 
\begin{equation}
\begin{split}
	&\delta^{3}(P)J_{p}dl_{1}dl_{2}dl_{3}\bigwedge\mathrm{dlog}\frac{\left\langle Y12\right\rangle }{\left\langle Y23\right\rangle }\\=&\frac{\delta^{3}\left(P\right)\left\langle Y13\right\rangle \left\langle 1234\right\rangle ^{2}}{\left\langle Y12\right\rangle \left\langle Y23\right\rangle }\left\langle Yd^{2}Y_{1}\right\rangle \left\langle Yd^{2}Y_{2}\right\rangle 
\end{split}
	\end{equation}
By extracting the functions in front of the measure, the function so called volume function correspond to the reduced SUSY amplitude in ABJM theory. Integrating out the Grassmann variables, the bracket $\langle1234\rangle$ corresponds to the super-momentum conservation 
\begin{equation}
	\left\langle 1234\right\rangle ^{2}\rightarrow\delta^{\left(4\right)}\left(q\right)
\end{equation}

\paragraph{$n=6$ case} The six-point case also only contains a single top cell. We can use the similar procedure to push-forward the top cell to $Y$ space. Here, we also choose our cell in the cyclic gauge
\begin{equation}
	\left(\begin{array}{cccccc}
c_{21} & 1 & c_{23} & 0 & c_{25} & 0\\
c_{41} & 0 & c_{43} & 1 & c_{45} & 0\\
c_{61} & 0 & c_{63} & 0 & c_{65} & 1
\end{array}\right)
\end{equation}
equation~\eqref{eq:ABJMhedron OG} and orthogonality condition can determine the matrix element
\begin{equation}\label{eq:OG36}
	c_{\bar{r}s}=\frac{\left\langle Y\bar{r}-2\,\bar{r}+2\right\rangle \left\langle Ys-2\,s+2\right\rangle -\sum_{i=1,3,5}\left\langle Y\bar{r}i\right\rangle \left\langle Ysi\right\rangle }{P_{135}^{2}}
\end{equation}
here the barred (unbarred) indices label even (odd)-particles, respectively.

Similar to the $n=4$ case, writing down  $\delta^3(P)d^3P$ require introduce regulators $l_1,l_2,l_3$ and we introduce them in this way
\begin{equation}
	\left(\begin{array}{cccccc}
l_{1}c_{21} & 1 & c_{23} & 0 & c_{25} & 0\\
c_{41} & 0 & l_{2}c_{43} & 1 & c_{45} & 0\\
c_{61} & 0 & c_{63} & 0 & l_{3}c_{65} & 1
\end{array}\right)
\end{equation}
Then the volume form is
\begin{equation}\label{eq:6ptform}
	\frac{\delta^{3}\left(P\right)\left(\sum_{i,j=1,3,5}\left\langle Yij\right\rangle \left\langle ij246\right\rangle +\left(1,3,5\right)\leftrightarrow\left(2,4,6\right)\right)^{2}P_{135}^{2}}{c_{25}c_{41}c_{63}}\left\langle Yd^{2}Y_{1}\right\rangle \left\langle Yd^{2}Y_{2}\right\rangle \left\langle Yd^{2}Y_{3}\right\rangle 
\end{equation}
where
\begin{equation}
 c_{\bar{r}s}= \left\langle Y\bar{r}-2\,\bar{r}+2\right\rangle \left\langle Ys-2\,s+2\right\rangle-\sum_{i=1,3,5}\left\langle Y\bar{r}i\right\rangle \left\langle Ysi\right\rangle 
\end{equation}
By removing the measure in $Y$ space, the extracted volume function can directly related the reduced SUSY amplitude in ABJM theory. All brackets in the denominator contain $\langle Y i j\rangle$ can directly translate to $\langle i j\rangle$, while the numerator require a more analysis. The reader can convince oneself that the numerator will reduce to $d^6 \lambda$ part:

\begin{equation}
	\left(\sum_{i,j=1,3,5}\left\langle Yij\right\rangle \left\langle ij246\right\rangle +\left(1,3,5\right)\leftrightarrow\left(2,4,6\right)\right)^{2}\rightarrow\delta^{\left(4\right)}\left(q\right)\left(\sum_{\bar{p},\bar{q},\bar{r}=2,4,6}\left\langle \bar{p}\bar{q}\right\rangle \eta_{\bar{r}}+\sum_{p,q,r=1,3,5}\left\langle pq\right\rangle \eta_{r}\right)^{2}
\end{equation}

As reviewed in appendix A, naively we have two branches when solving orthogonal conditions, and for $n=6$ both of them naively contribute to BCFW pushforward. What we have shown above is the contribution from positive branch, and naively there is a canonical form from  the negative branch. It takes a very similar form as \eqref{eq:6ptform}, but in the numerator we need to change the plus sign to a minus one. It turns out that the numerator vanishes by momentum conservation, thus there is no contribution to the form from negative branch. 

\subsection{The $D=3$ twistor-string map and the pushforward}

Let us first define the Veronese map from $\mathcal{M}_{0,n}^+$ to orthogonal Grassmannian OG$(k,n=2k)$:
\begin{equation}\label{eq:veronese OG}
	\left(\begin{array}{cccc}
1 & 1 & \cdots & 1\\
\sigma_{1} & \sigma_{2} & \cdots & \sigma_{n}
\end{array}\right)\longrightarrow C_{\alpha i}(\sigma)=\left(\begin{array}{ccccc}
t_{1} & t_{2} & \cdots & t_{n-1} & 1\\
t_{1}\sigma_{1} & t_{2}\sigma_{2} & \cdots & t_{n-1}\sigma_{n-1} & \sigma_{n}\\
\\
t_{1}\sigma_{1}^{k-1} & t_{2}\sigma_{2}^{k-1} & \cdots & t_{n-1}\sigma_{n-1}^{k-1} & \sigma_{n}^{k-1}
\end{array}\right)
\end{equation}
where $t^2_{i}=(-1)^i\frac{\prod_{j\neq n}\sigma_{nj}}{\prod_{j\neq i}\sigma_{ij}}=(-1)^i\frac{v_n}{v_i}$ for $i=1,\dots,n-1$ (with $t_n=1$). To see this, we start with the Veronese map from $G(2,n)$ to $G(k, n)$ with $k=n/2$, and impose orthogonal conditions $C \cdot \Omega \cdot C^T=0$ which translate into $n{-}1$ equations of the form $\sum_{i=1}^n (-1)^{i-1} t_i^2 \sigma_i^{\alpha}=0$ with $\alpha=0,1,\cdots,n-2$; using $GL(1)$ to fix $t_n=1$ we can obtain solutions for $t_i^2$ as above. 

Again with positive bosonized kinematic variables, $\Lambda^A_i$, which additionally satisfy the conditions that all planar variables are positive, we can define our $D=3$ twistor-string map as $\Phi_{\Lambda}: \mathcal{M}_{0,n}^+\to G(k,k{+}2)$
via \begin{equation}\label{eq:3dts}
	Y_{\alpha}^{A}=\sum_{i=1}^n C_{\alpha i}\left(\sigma\right)\,\Lambda_{i}^{A}
\end{equation}
with $C_{\alpha i}(\sigma)$ in~\eqref{eq:veronese OG}. Alternatively, we can directly map to the $(n{-}3)$-dim subspace spanned by $y$ (with fixed $\Delta, \lambda^*$, see \eqref{eq:fix3d}), where the map are simply $D=3$ scattering equations ($2k{-}3$ of them independent) 
\begin{equation}\label{eq:3dts2}
C\left(\sigma\right) \cdot \lambda (y)=0,
\end{equation}
In both forms, note that from orthogonality conditions there are two sign choices for each $t_i=\pm \sqrt{(-1)^i v_n/v_i}$ ($i=1,\cdots, n{-}1$), only one of them satisfies $D=3$ scattering equations, thus the $t$'s are completely determined by $\sigma$'s for each of the $E_{2k{-}3}$ solutions. This is why the moduli space is basically ${\cal M}_{0,n}^+$ (instead of $G_+(2,n)$). 

We conjecture that the image $\Phi_{\Lambda}({\cal M}_{0,n}^+)$ gives the momentum amplituhedron of ABJM, ${\cal M}^{\rm 3d}(k,2k)$, and the map is a diffeomorphism from ${\cal M}_{0,n}^+$ to  its interior. Similar to the $D=4$ case, any point in ${\cal M}_{0,n}^+$ clearly maps to a point in ${\cal M}^{\rm 3d}(k,2k)$, so the non-trivial part of the conjecture is if the map is onto and one-one. We have performed thorough checks with many kinematic data points generated by BCFW cells up to $n=2k=8$: in all cases we find that given a point in ${\cal M}^{\rm 3d}(k,2k)$, it turns out that out of $E_{2k{-}3}$ solutions of $D=3$ scattering equations, there is a unique {\it positive solution} with $\sigma_i<\sigma_j$ for $i<j$ (thus it is in ${\cal M}_{0,n}^+$) and $t_i>0$! Note that a prior we do not care about $t_i$ for the pre-image to be in ${\cal M}_{0,n}^+$, but for the image to be in $OG_+(k,2k)$ we need positive $t_i$ as well. 

In fact, this unique positive solution in $D=3$ is exactly the same one for the middle sector $k=n/2$ in $D=4$, if we identify 4d data from 3d data. Given any $D=3$ kinematic data $\{\lambda_i^{(3)}\}$ with positivity and correct sign flip, we identify it with a  special kinematic point in $D=4$ via $$    \lambda^{(4)}_i=\lambda^{(3)}_i\cdot\Omega\qquad\tilde{\lambda}^{(4)}_i=\lambda^{(3)}_i,$$
and clearly the scattering equations and momentum conversation in $D=4$ are still satisfied. If our conjecture in $D=4$ holds, which means for this special kinematic point we have a unique positive solution, we immediately see that it is a positive solution for $D=3$ (again the $t_i$'s are determined by the $\sigma$'s). We have checked explicitly for the highly non-trivial case with $n=2k=8$: given $D=3$ data we find that the unique positive solution of $D=4$ scattering equations indeed agrees with that of $D=3$. Note that other $D=4$ solutions do not trivially reduce to $D=3$ ones in each sector. As already shown in~\cite{Cachazo:2013iaa}, with $D=3$ $n=8$ kinematics: out of $5!=120$ solutions of CHY scattering equations, there are $16$ solutions without multiplicity (in $k=4$ sector), $22$ solutions with multiplicity $4$ (in $k=3$ sector) and $1$ solution with multiplicity $16$ (in $k=2$ sector).

In the next subsection, we will also check how to map boundaries of ${\cal M}_{0,n}^+$ to those of  ${\cal M}^{\rm 3d}(k,2k)$, which in particular shows that our $D=3$ twistor-string map indeed knows about the interesting factorizations of ABJM amplitudes. In the rest of this subsection, we study the pushforward from ${\cal M}_{0,n}^+)$ to the canonical form of ${\cal M}^{\rm 3d}(k,2k)$, which gives ABJM amplitudes with reduced SUSY. The derivation is similar to that in~\cite{Cachazo:2013iaa}, and we also fix the normalization constant of this pushforward.

Recall that with reduced SUSY, the twistor-string formula becomes
\begin{equation}\label{eq:rudsusy}
	A_{n}^{\mathrm{3d, reduced}}=\frac{1}{\mathrm{vol}\,GL\left(2\right)}\int\prod_{i}^{odd}d\eta_{i}^{3}\int d^{2n}C\frac{J\Delta\prod_{\alpha=1}^{k}\delta^{2|3}\left(C\cdot\Lambda^{\mathrm{}}\right)}{\left(12\right)\left(23\right)\cdots\left(n1\right)},
\end{equation}
and what we will show is that it can be rewrite as 
\begin{equation}\label{eq:cans3d}
	A_{n}^{\mathrm{3d, reduced}}=\frac{1}{\mathrm{vol}\,SL\left(2\right)}~\int d^{n}\sigma_{i}\frac{\delta^{2|2}\left(C\cdot\Lambda^{\mathrm{reduced}}\right)}{\sigma_{12}\sigma_{23}\cdots\sigma_{n1}}
\end{equation}
where $\Lambda^{\mathrm{reduced}}=\left(\left|i\right\rangle ,\eta_{i}^{A}\right), A=1,2$ and in the Veronese $C$ matrix we plug in the solution for $t_i$ with correct sign, for each solution of $\sigma$'s. 

First, we change the variables to inhomogeneous coordinates, $\left(a_{i},b_{i}\right)=t_{i}^{\frac{1}{d}}\left(1,\sigma_{i}\right)$ (recall $d=k{-}1$). This results in a Jacobian factor 
$\frac{1}{d^{n}}\prod_{i=1}^{n}t_{i}^{{-}1{+}\frac{2}{d}}$, and the entries of the matrix become $C_{\alpha i}=t_{i}\sigma_{i}^{\alpha{-}1}$ with $\Delta=\prod_{\alpha=0}^{2d}\delta\left(\sum_{i}(-1)^{i-1}t_{i}^{2}\sigma_{i}^{\alpha}\right)$, as well as
\begin{equation}
	J=\frac{\prod_{1\leq i<j\leq2d{+}1}t_{i}^{\frac{1}{d}}t_{j}^{\frac{1}{d}}\sigma_{ij}}{\prod_{1\leq i<j\leq d{+}1}t_{2i{-}1}^{\frac{1}{d}}t_{2j{-}1}^{\frac{1}{d}}\sigma_{2i{-}1,2j{-}1}}, \quad \frac{1}{\left(12\right)\left(23\right)\cdots\left(n1\right)}=\prod_{i=1}^{n}t_{i}^{{-}\frac{2}{d}}\frac{1}{\sigma_{12}\sigma_{23}\cdots\sigma_{n1}}
\end{equation}

Next, localizing $\Delta$ by integrating out $t_{i}$ for $i=1,\cdots,n{-}1$ (with $t_n=1$) gives  $t_{i}=\pm \sqrt{(-1)^i v_n/v_i}$ with the sign fixed by $C \cdot \Lambda=0$; it also picks up another Jacobian factor $\frac{1}{2^{n{-}1}}\prod_{i=1}^{n{-}1} t_i v_i$. Finally, we integrate out $\eta_{i}$ for all the odd $i$'s, which gives another factor $\prod_{i=1}^{d{+}1}t_{2i{-}1}\prod_{1\leq i<j<d{+}1}\sigma_{2i{-}1, 2j{-}1}$. Collecting all factors together, we arrive at \eqref{eq:rudsusy}; by $\eta_i \to d \lambda_i$, we obtain the pushforward formula for the canonical form:
\begin{equation}\label{eq:3dcf}
	\bold{\Omega}({\cal M}^{\rm 3d}(k,2k))=\sum_{\rm sol.}^{E_{n{-}3}} \frac{d^{n{-}3} \sigma}{c_n~\sigma_{12}\sigma_{23}\cdots\sigma_{n1}}
\end{equation}
where $c_n=2^{n{-}1} d^n=(n-2)^n/2$ is a normalization factor: our pushforward uses $\frac 1 {c_n} \Omega({\cal M}_{0,n}^+)$ which can be interpreted as defining residues of $0$-dim boundaries (points) as $\pm 1/c_n$. For $n=2k=4,6,8$, we have also checked numerically that the pushforward with correct normalization \eqref{eq:cans3d} indeed gives $\Omega({\cal M}^{\rm 3d}(k,2k))$ as obtained using BCFW cells above, where we need to sum over $1, 2$ and $16$ solutions respectively. 

\subsection{Boundaries of ${\cal M}^{\rm 3d}(k,2k)$}
In this subsection, we study the very interesting problem of how the boundaries of compactified $\mathcal{\overline{M}}_{0,n}^+$ get mapped to those of $\mathcal{M}^{\text{3d}}(k,2k)$ under the $D=3$ twistor-string map. Unlike SYM amplitudes, ABJM amplitudes only have multi-particle poles which correspond to factorizing into two even-point amplitudes,  except for the special two-particle poles for $n=4$~\cite{Huang:2011um}. Therefore, for $n=4$, $\mathcal{M}_{k,2k}^{\text{3d}}$ has (co-dimensional one) boundaries :
\begin{equation}
	\left\langle Y12\right\rangle =\left\langle Y34\right\rangle \quad\mathrm{and}\quad\left\langle Y23\right\rangle =\left\langle Y14\right\rangle 
\end{equation}
while for $n>4$, only planar Mandelstam  variables with odd number of labels are co-dimension one boundaries
\begin{equation}
	S_{i,i+1,\cdots,j}=\sum_{i\leq p<q\leq j}\left(-1\right)^{p+q+1}\left\langle Ypq\right\rangle ^{2},\quad\left|j-i\right|=\mathrm{even}.
\end{equation}
It is important that those with even number of labels including collinear poles, which we call unwanted channels, are no longer co-dimensional one boundaries. Remarkably, we will provide strong evidence that our $D=3$ twistor-string map knows about this intriguing pattern of boundaries of ${\cal M}^{\rm 3d}(k,2k)$, {\it i.e.} it automatically exclude those unwanted channels.

We will first study the special $n=4$ case and then proceed to $n=6,8, 10$ which are generic enough. We will exhaust all possible boundaries of $\mathcal{\overline{M}}_{0,n}^+$ and we observe that in a rather beautiful way, our $D=3$ map does distinguish these channels as it pushes non-factorization ones to higher co-dimension boundaries of ${\cal M}^{\rm 3d}(k,2k)$. 

\paragraph{$n=4$ case} As it is familiar from our $D=4$ discussions, we compactify $\mathcal{M}_{0,n}^+$ by introducing $u$ variables. After solving $t$'s, the Veronese map~\eqref{eq:veronese OG} for $n=4$ becomes nicely
\begin{equation}
	C_{2\times 4}=\left(\begin{array}{cccc}
\sqrt{u_{2,4}} & 1 & \sqrt{u_{1,3}} & 0\\
-\sqrt{u_{1,3}} & 0 & \sqrt{u_{2,4}} & 1
\end{array}\right),
\end{equation}
with $u_{1,3}+u_{2,4}=1$. In this case, we only have two boundaries $u_{1,3}\to 0$ and $u_{2,4}\to 0$, which correspond to two-particle poles (special for $n=4$). Similar to MHV case in $D=4$, $C_{2\times4}$ is related to $(\lambda_i)_{2\times4}$ by a GL(2), thus as $u_{1,3} \to 0$ (first two columns proportional), we have $\langle Y12\rangle=\langle Y34\rangle=0$ and similarly as $u_{2,4}\to 0$ we have $\langle Y14\rangle=\langle Y23\rangle=0$. In each case, we find a co-dimension one boundary of this (one-dimensional) geometry, ${\cal M}^{\rm 3d}(2,4)$.

\paragraph{$n>4$ cases} For general $n$, we use $u$ variables to compactify $\mathcal{M}_{0,n}^+$ and study the image of all boundaries as $u_{i,j}\to 0$ under the $D=3$ map. For example the Veronese map~\eqref{eq:veronese OG} in terms of $u$'s for $n=6$ reads:
\begin{equation}
	C_{3\times6}=\left(\begin{array}{cccccc}
\sqrt{u_{2,4}u_{2,6}} & 1 & \sqrt{u_{1,3}u_{3,5}} & 0 & -\sqrt{u_{1,3}u_{1,4}^{2}u_{1,5}u_{2,4}u_{4,6}} & 0\\
-\sqrt{u_{1,3}u_{2,6}u_{3,5}u_{3,6}^{2}u_{4,6}} & 0 & \sqrt{u_{2,4}u_{4,6}} & 1 & \sqrt{u_{1,5}u_{3,5}} & 0\\
\sqrt{u_{1,3}u_{1,5}} & 0 & -\sqrt{u_{1,5}u_{2,4}u_{2,5}^{2}u_{2,6}u_{3,5}} & 0 & \sqrt{u_{2,6}u_{4,6}} & 1
\end{array}\right).
\end{equation}
Something very interesting already happens here with collinear poles. If we send $u_{1,3}\to 0$ which forces $u_{2,4}=u_{2,5}=u_{2,6}=1$, it is easy to see that the $C_{3\times 6}$ matrix becomes {\it block diagonal} with a trivial $C^L_{1\times 2}=(1,1)$ upper-left block and a $C^R_{2\times 4}$ lower-right block, which resembles the $n=4$ matrix above. This form makes it clear that we arrive at a higher co-dimensional boundary; in particular since the first two columns are both $(1, 0, 0)^{\rm T}$), we immediately see that $\lambda_1=\lambda_2$, {\it i.e.} they are not only proportional to each other but actually {\it identical} ! In addition to $S_{12}=0$, we have that for any $a=3,\cdots, n$:
$$
S_{1,2,a}=\langle Y 1 2\rangle^2+(-)^a (\langle Y 1 a\rangle^2-\langle Y 2 a\rangle^2)=0
$$
thus $u_{1,3}\to 0$ is not mapped to a co-dimension one boundary for $n=6$. In general, one can show that for $n>4$, $u_{i,i+2}\to 0$ is no longer mapped to a co-dimension one boundary since we have $\lambda_i=\lambda_{i{+}1}$ and consequently $S_{i,i+1,j}$ for any $j$ (in particular $j=i{+}2$ or $i{-}1$) vanish, in addition to $S_{i,i{+}1}=0$.

On the other hand, if we send $u_{1,4}\to 0$ (which forces $u_{2,5}=u_{2,6}=u_{3,5}=u_{3,6}=1$), we see that the $C_{3\times 6}$ matrix degenerates to co-dimension one; this leads to a co-dimension one boundary with $S_{123}=0$ and no other planar variables vanish. This clearly illustrate the difference between these two types of channels already for $n=6$.

Next we move to $n=8$ and $n=10$; first we have the $C_{4\times 8}$ matrix for $n=8$ as
\begin{equation}
	C_{4\times8}=\left(\begin{array}{cccccccc}
c_{1}\left(1\right) & 1 & c_{4}\left(3\right) & 0 & -c_{3}\left(5\right) & 0 & c_{2}\left(7\right) & 0\\
-c_{2}\left(1\right) & 0 & c_{1}\left(3\right) & 1 & c_{4}\left(5\right) & 0 & -c_{3}\left(7\right) & 0\\
c_{3}\left(1\right) & 0 & -c_{2}\left(3\right) & 0 & c_{1}\left(5\right) & 1 & c_{4}\left(7\right) & 0\\
-c_{4}\left(1\right) & 0 & c_{3}\left(3\right) & 0 & -c_{2}\left(5\right) & 0 & c_{1}\left(7\right) & 1
\end{array}\right)
\end{equation}
where we have defined 
\begin{equation*}
	\begin{split}
		&c_{1}\left(i\right):=\sqrt{u_{i+1,i+3}u_{i+1,i+5}u_{i+1,i+7}},
		\quad c_{4}\left(i\right):=\sqrt{u_{i+2,i+8}u_{i+4,i+8}u_{i+6,i+8}}\\
		&c_{2}\left(i\right):=\sqrt{u_{i+1,i+5}u_{i+1,i+7}u_{i+2,i+4}u_{i+2,i+5}^{2}u_{i+2,i+6}u_{i+2,i+7}^{2}u_{i+2,i+8}u_{i+3,i+5}u_{i+3,i+7}}\\
		&c_{3}\left(i\right):=\sqrt{u_{i+1,i+7}u_{i+2,i+6}u_{i+2,i+7}^{2}u_{i+2,i+8}u_{i+3,i+7}u_{i+4,i+6}u_{i+4,i+8}u_{i+4,i+7}^{2}u_{i+5,i+7}},
	\end{split}
\end{equation*}
and similarly the $C_{5\times 10}$ matrix for $n=10$:

\begin{equation}
	C_{5\times10}=\left(\begin{array}{cccccccccc}
c_{1}\left(1\right) & 1 & c_{5}\left(3\right) & 0 & -c_{4}\left(5\right) & 0 & c_{3}\left(7\right) & 0 & -c_{2}\left(9\right) & 0\\
-c_{2}\left(1\right) & 0 & c_{1}\left(3\right) & 1 & c_{5}\left(5\right) & 0 & -c_{4}\left(7\right) & 0 & c_{3}\left(9\right) & 0\\
c_{3}\left(1\right) & 0 & -c_{2}\left(3\right) & 0 & c_{1}\left(5\right) & 1 & c_{5}\left(7\right) & 0 & -c_{4}\left(9\right) & 0\\
-c_{4}\left(1\right) & 0 & c_{3}\left(3\right) & 0 & -c_{2}\left(5\right) & 0 & c_{1}\left(7\right) & 1 & c_{5}\left(9\right) & 0\\
c_{5}\left(1\right) & 0 & -c_{4}\left(3\right) & 0 & c_{3}\left(5\right) & 0 & -c_{2}\left(7\right) & 0 & c_{1}\left(9\right) & 1
\end{array}\right)
\end{equation}
where we have defined (not to be confused with those entries in $n=8$ case):
\begin{equation*}
	\begin{split}
		&c_{1}\left(i\right):=\sqrt{u_{i+1,i+3}u_{i+1,i+5}u_{i+1,i+7}u_{i+1,i+9}},\ c_{5}\left(i\right):=\sqrt{u_{i+2,i+10}u_{i+4,i+10}u_{i+6,i+10}u_{i+8,i+10}}
	\end{split}
\end{equation*}
\begin{equation*}
	\begin{split}
				&c_{2}\left(i\right):=\left(u_{i+1,i+5}u_{i+1,i+7}u_{i+1,i+9}u_{i+2,i+4}u_{i+2,i+5}^{2}u_{i+2,i+6}u_{i+2,i+7}^{2}\right.\\
		&\qquad \qquad \quad \left. u_{i+2,i+8}u_{i+2,i+9}^{2}u_{i+2,i+10}u_{i+3,i+5}u_{i+3,i+7}u_{i+3,i+9}^{\,}\right)^{1/2}\\
		&c_{3}\left(i\right):=\left(u_{i+1,i+7}u_{i+1,i+9}u_{i+2,i+6}u_{i+2,i+7}^{2}u_{i+2,i+8}u_{i+2,i+9}^{2}u_{i+2,i+10}u_{i+3,i+7}\right.\\
		&\qquad \qquad \left.u_{i+3,i+9}u_{i+4,i+6}u_{i+4,i+7}^{2}u_{i+4,i+8}u_{i+4,i+9}^{2}u_{i+4,i+10}u_{i+5,i+7}u_{i+5,i+9}\right)^{1/2}\\
		&c_{4}\left(i\right):=\left(u_{i+1i+,9}u_{i+2,i+8}u_{i+2,i+9}^{2}u_{i+2,i+10}u_{i+3,i+9}u_{i+4,i+8}u_{i+4,i+9}^{2}\right.\\
		&\qquad \qquad \quad \left.u_{i+4,i+10}u_{i+5,i+9}u_{i+6,i+8}u_{i+6,i+9}^{2}u_{i+6,i+10}u_{i+7,i+9}\right)^{1/2}.
	\end{split}
\end{equation*}
It is straightforward to write such $u$ parametrization of $C_{k\times 2k}$ matrix for any $n=2k$ points. 

We first study four-particle poles, {\it e.g.} for $n=8$, the boundary $S_{1234}=0$ can be realized by sending $u_{1,5}\to 0$ which forces $u_{2,6}=u_{2,7}=u_{2,8}=u_{3,6}=u_{3,7}=u_{3,8}=u_{4,6}=u_{4,7}=u_{4,8}=1$. By plugging these into $C_{4\times 8}$ above, remarkably it becomes block diagonal with two $2\times 4$ blocks as follows (which is again highly degenerate):
$$\left(
\begin{array}{cccccccc}
 \sqrt{u_{2,4}} & 1 & \sqrt{u_{1,3} u_{3,5}} & 0 & 0 & 0 & 0 & 0 \\
 -\sqrt{u_{1,3} u_{3,5}} & 0 & \sqrt{u_{2,4}} & 1 & 0 & 0 & 0 & 0 \\
 0 & 0 & 0 & 0 & \sqrt{u_{6,8}} & 1 & \sqrt{u_{1,7} u_{5,7}} & 0 \\
 0 & 0 & 0 & 0 & -\sqrt{u_{1,7} u_{5,7}} & 0 & \sqrt{u_{6,8}} & 1 \\
\end{array}
\right).$$
Note that the both $C_{2\times 4}$ blocks look very similar to $n=4$ case. In fact, the $u$ equations for remaining variables reduce to $1-u_{2,4}=u_{1,3} u_{3,5}$ and $1-u_{6,8}=u_{1,7} u_{5,7}$; it is convenient to redefine RHS as $u_{1,3} u_{3,5}=u'_{1,3}$ and $u_{1,7} u_{5,7}=u'_{5,7}$, then we see literally two $n=4$ blocks. Therefore, $D=3$ equations $C \cdot \lambda=0$ become decoupled into two sets of equations, one for particles $1,2,3,4$ and the other one for $5,6,7,8$. In particular we see that the kinematics become very degenerate under the map, since we have $p_1+p_2+p_3+p_4=0$ and $p_5+p_6+p_7+p_8=0$. This tells us that in addition to $S_{1234}=0$ we have $S_{123}=S_{124}=S_{134}=S_{234}=0$, {\it i.e.} all three-particle variables in $1,2,3,4$ (same for $5,6,7,8$) vanish as well! Thus we see that $S_{1234}=0$ is a higher co-dimension boundary as expected.  

Exactly the same reason exclude all such four-particle poles as co-dimension one boundaries for higher $n$; {\it e.g.} for $n=10$ with $u_{1,5}\to 0$, the $C_{5\times 10}$ matrix become block-diagonal with $C^L_{2\times 4}$ and $C^R_{3\times 6}$. In addition to $S_{1234}=0$ we find all three-particle variables with labels in $1,2,3,4$ and all five-particle variables in $5,6,7,8,9,10$ vanish. We have checked for $n=12$ and excluded six-particle poles as co-dimension one boundaries as well.  

We also study {\it e.g.} $u_{1,4}\to 0$ for $n=8,10$, where, similar to $n=6$ case, the matrix only degenerates to co-dimension one and we have only $S_{123}=0$, which is a co-dimension one boundary. Finally, we look at five-particle poles for $n=10$: {\it e.g.} with $u_{16}\to 0$, we find that $C_{5\times 10}$ degenerates to co-dimension one, and $S_{12345}=0$ is beautifully a co-dimension one boundary with no other planar variables vanish.

To conclude, we have very strong evidence that, under the $D=3$ map, any unwanted (odd-odd) channel (or planar Mandelstam pole with even number of labels) is not a co-dimension one boundary since it forces other planar variables to vanish. On the contrary, all planar poles with odd number of labels (or even-even factorization channels), do correspond to co-dimension one boundaries. We expect a simple proof of this claim to all multiplicities. 

We also compare boundary structures of momentum amplituhedra of SYM and ABJM. We have discussed boundaries for SYM case above, but we wanted to compare the $C$ matrices with those in ABJM directly. For example, for the middle sector $k=4$, $n=8$, the $D=4$ Veronese map~\eqref{eq:CV} from $G_+(2,n)$ to $G_+(k,n)$ reads:
\begin{equation}
	C_{4\times8}=\left(\begin{array}{cccccccc}
c_{1}\left(1\right) & 1 & c_{4}\left(3\right) & 0 & -c_{3}\left(5\right) & 0 & c_{2}\left(7\right) & 0\\
-c_{2}\left(1\right) & 0 & c_{1}\left(3\right) & 1 & c_{4}\left(5\right) & 0 & -c_{3}\left(7\right) & 0\\
c_{3}\left(1\right) & 0 & -c_{2}\left(3\right) & 0 & c_{1}\left(5\right) & 1 & c_{4}\left(7\right) & 0\\
-c_{4}\left(1\right) & 0 & c_{3}\left(3\right) & 0 & -c_{2}\left(5\right) & 0 & c_{1}\left(7\right) & 1
\end{array}\right),
\end{equation}
where (again not to be confused with entries above)
\begin{equation*}
	\begin{split}
		&c_{1}\left(i\right):=\frac{t_i}{t_{i+1}}\sqrt{u_{i+1,i+3}u_{i+1,i+5}u_{i+1,i+7}},
		\quad c_{4}\left(i\right):=\frac{t_i}{t_{i+7}}\sqrt{u_{i+2,i+8}u_{i+4,i+8}u_{i+6,i+8}}\\
		&c_{2}\left(i\right):=\frac{t_i}{t_{i+3}}\sqrt{u_{i+1,i+5}u_{i+1,i+7}u_{i+2,i+4}u_{i+2,i+5}^{2}u_{i+2,i+6}u_{i+2,i+7}^{2}u_{i+2,i+8}u_{i+3,i+5}u_{i+3,i+7}}\\
		&c_{3}\left(i\right):=\frac{t_i}{t_{i+5}}\sqrt{u_{i+1,i+7}u_{i+2,i+6}u_{i+2,i+7}^{2}u_{i+2,i+8}u_{i+3,i+7}u_{i+4,i+6}u_{i+4,i+8}u_{i+4,i+7}^{2}u_{i+5,i+7}}.
	\end{split}
\end{equation*}
Note that the only difference with $D=3$ case is it has $t_i$ variables which can scale independently ($D=3$ case is obtained by setting all $t_i=1$). We see that the $t_i$ variables not only give collinear singularities in $D=4$ but they are also responsible for the appearance of odd-odd factorizations for SYM. For example, here $S_{1234}=0$ is a co-dimensional one boundary: as we send $u_{1,5} \to 0$ and scale $t_1, t_2, t_3, t_4$ as $1/\sqrt{u_{1,5}}$ (or $t_5, t_6, t_7, t_8$ as $\sqrt{u_{1,5}}$), the $C$ matrix only degenerates to co-dimension one and we find only $S_{1234}=0$ (no other planar variables vanish). In general, with appropriate scaling for $t$'s, all factorization channels exist as co-dimension one boundaries for SYM amplituhedron.

\section{Conclusion and Discussions}

In this paper we propose twistor-string maps in $D=4$ and $D=3$ from moduli space to the momentum amplituhedron for SYM and ABJM respectively. The latter is a new positive geometry defined in terms of $OG_+(k, 2k)$, similar to that of SYM in terms of $G_+(k,n)$, and we find a common origin for both geometries via the pullback of scattering equations to $(2n{-}4)$ or $(n{-}3)$-dim subspace in their kinematic space. There are numerous open questions raised by our preliminary studies. For example, our ABJM momentum amplituhedron is in some sense a ``dimension reduction" of the SYM one, and could we find the long sought-after amplituhedron in 3d momentum-twistor space~\cite{Elvang:2014fja} for ABJM (even at tree level) in a similar way? Moreover, are there similar twistor-string maps for momentum amplituhedron with other $m$ such as $m=2$~\cite{Lukowski:2020dpn,Parisi:2021oql}?

One of the most pressing tasks is to find a rigorous proof of our main conjectures both for $D=4$ and $D=3$, which among other things would provide two remarkable examples of diffeomorphism between positive geometries beyond the well-understood polytope case~\cite{Arkani-Hamed:2017mur}. Moreover, a fascinating property of the maps is that while ${\cal M}_{0,n}^+$ has the same boundary structures after compactifications, our $D=4$ and $D=3$ maps know precisely boundary structures of corresponding momentum amplituhedra: for the former we have additionally soft-collinear singularities related to $t$ variables, and for the latter all unwanted channels (odd-odd factorizations), except for collinear ones for $n=4$, are automatically excluded. It would be highly desirable to understand all these better.  

\begin{figure}[H]
    \centering
    \includegraphics[scale=0.8]{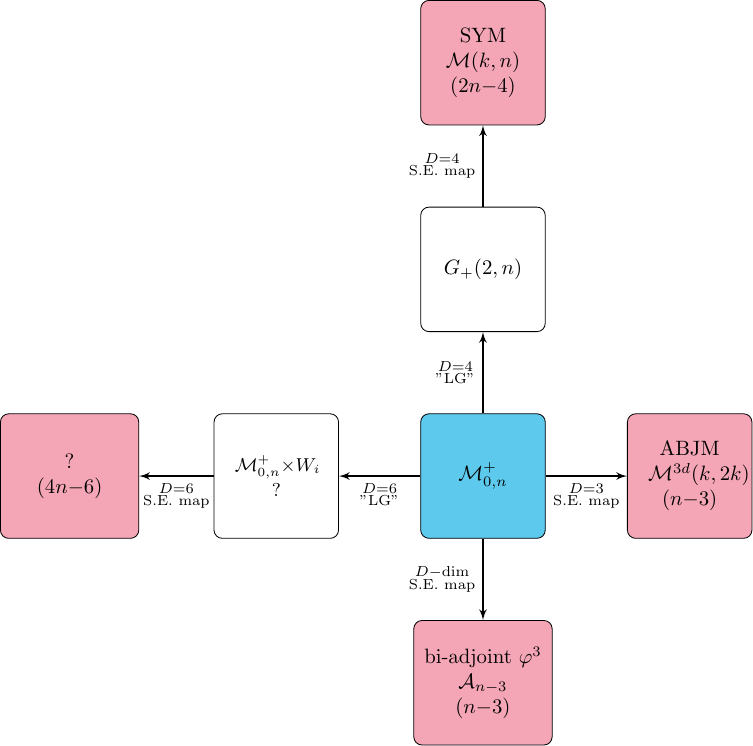}
\caption{Maps from moduli spaces to ``amplituhedra" in general $D$ and $D=3,4$ {\it etc.}}
    \label{fig:monm}
\end{figure}

Our construction provides a unified picture for ``kinematic amplituhedra" including momentum amplituhedra for SYM and ABJM, and the associahedron for bi-adjoint $\phi^3$ theory. All three originate from the same worldsheet associahedron ${\cal M}_{0,n}^+$ but with different maps and ``target spaces". For $\phi^3$ in general dimension, the CHY scattering equations can be used to map it to ${\cal A}_{n{-}3}$ in Mandelstam space. For kinematic space in specific dimension, we need to ``uplift" the moduli space to include ``little group” (LG) redundancy: for $D=3$, it is $\mathbb{Z}_2$ and the moduli space is still  ${\cal M}_{0,n}^+$ (up to a normalization constant), but in $D=4$ we have $U(1)$ (or $GL(1)$ for complexified kinematics) and the moduli space become $G_+(2,n)$; then via twistor-string maps (or scattering equations) in $D=4$ and $D=3$ we have the corresponding amplituhedron ${\cal M}(k,n)$ and ${\cal M}^{\rm 3d}(k,2k)$ respectively, as shown in figure~\ref{fig:monm}.

A very natural question is if we can extend our construction to other dimensions, such as $D=6$ where twistor-string-like formulas have been studied extensively~\cite{Cachazo:2018hqa, Schwarz:2019aat, Geyer:2018xgb}. The little group is $SL(2)$ for complexified kinematics (or $SU(2)$ for real case), and we expect a moduli space of dimension $n{-}3+3(n{-}1)=4n{-}6$~\footnote{Note that the dimension of the space is $(D{-}2)n{-}D$ for $D=3,4,6$, which is expected to hold for some other $D$ as well.}. The analog of $t_i$ (and $\tilde{t}_i=1/(t_i v_i)$) variables are the $2\times 2$ matrix $W_i$ of~\cite{Schwarz:2019aat}, which satisfy the constraint $\det W_i=1/v_i$ (thus only $3$ degrees of freedom each). With a possible notion of positivity for the $W_i$'s, one may define the positive part of this moduli space ${\cal M}_{0,n} \times \{ W_i| \det W_i=1/v_i\}/{\rm SL}(2)$, and even attempt to construct a $D=6$ twistor-string map through a Veronese map to the symplectic Grassmannian $LG_+(n, 2n)$~\cite{Arkani-Hamed:2012zlh}. What is the resulting positive geometry (of dimension $4n{-}6$)? Could it be interpreted as the momentum amplituhedron for certain $D=6$ amplitudes? We leave these fascinating questions to future investigations. 

We have seen that twistor-string maps (or scattering equations in $D=4,3$) are powerful tools for studying not only tree amplitudes/canonical forms but also the underlying positive geometries including their boundaries. There are other intriguing features of the geometries and forms, which are made manifest by such maps. While we are limited to a given ordering in momentum twistor space, it is natural to consider relations among amplitudes with different orderings, such as KK (see~\cite{Damgaard:2021qbi}) and BCJ relations~\cite{Bern:2008qj}. It is well known that twistor-string formulas (or pushforward) trivialize KK relations, as well as BCJ relations on the support of scattering equations~\cite{Cachazo:2012da}. It would be interesting to see implications of these relations on the geometries via twistor-string maps. Moreover, while we do not have positive geometries for (super-)gravity amplitudes, their twistor-string formulas may provide insights towards a geometric picture, and even geometric understanding of double-copy~\cite{Bern:2008qj}, especially for SUGRA amplitudes from SYM and ABJM ones (see~\cite{Bargheer:2012gv, Huang:2012wr} for $D=3$ examples).

Recall that the kinematic associahedron naturally emerges as (the Minkowski sum of) Newton polytopes of string integrals, whose saddle-point equations provide the scattering equations and the corresponding map~\cite{Arkani-Hamed:2019mrd}. This idea generalizes to ``stringy canonical forms" for generic polytopes, where saddle-point equations always provide such a diffeomorphism~\cite{Arkani-Hamed:2019mrd, Arkani-Hamed:2017tmz}. Given our (non-polytopal) momentum amplituhedra in $D=3,4$, it is tempting to wonder if they have natural deformations into ``stringy" integrals, from which our twistor-string maps nicely emerge from the corresponding saddle-point equations. Relatedly, it would be interesting to find universal geometric/combinatorial interpretation for the number of solutions (or intersection number), namely $(n{-}3)!$ for general $D$ (including $D=6$), $E_{n{-}3, k{-}2}$ for $D=4$, and $E_{2k{-}3}$ for (the middle sector of) $D=3$. 

\section*{Acknowledgement}
It is a pleasure to thank Yu-tin Huang, Ryota Kojima, Congkao Wen, Shun-Qing Zhang for inspiring discussions. SH's research is supported in part by National Natural Science Foundation of China under Grant No. 11935013,11947301, 12047502,12047503. C.-K. K. is supported by MoST Grant No. 109-2112-M-002 -020 -MY33.

\appendix

\section{Grassmannian formulas and BCFW cells for SYM and ABJM}\label{review}
\subsection{SYM}
The Grassmannian formalism in $\mathcal{N}=4$ SYM originates from trivializing $D=4$ momentum conservation $\sum_{i=1}^n \lambda_i\tilde{\lambda}_i=0$ by introducing a $k$ plane in $n$ dimension $C\in G(k,n)$ such that
\begin{equation}
C^{\perp} \cdot \lambda=0 \quad C \cdot \tilde{\lambda}=0\,.
\end{equation}

This story then plays an important role in computing amplitudes in $\mathcal{N}=4$ SYM. To be specify, the $n$-pt $\mathrm{N^{k-2}MHV}$ tree amplitude can be written as contour integral around certain cells of $G_+(k,n)$~\cite{ArkaniHamed:2009dn,Arkani-Hamed:2012zlh}:

\begin{equation}\label{eq:gA}
\oint_{\gamma}\frac{d^{k \times n} C}{\mathrm{Vol\;(GL}(k))} \frac{\delta^{0 \mid 2 k}(C \cdot \tilde{\eta}) \delta^{0 \mid 2(n-k)}\left(C^{\perp} \cdot \eta\right)}{(1 \cdots k) \cdots(n \cdots k-1)} \delta^{2 k}(C \cdot \lambda) \delta^{2(n-k)}\left(C^{\perp} \cdot \tilde{\lambda}\right)\,.
\end{equation}

The contour $\gamma$ can be physically determined the BCFW recursion relations. And performing the integral~\eqref{eq:gA} means
evaluating a sum of residues, with each residue corresponding to a $2n{-}4$ dimension positroid cell in the positive Grassmannian $G_+(k,n)$.

For example, in the $4$pt  MHV case, we already in the top cell. After gauge fixing, the cell can be parametrize with $4$ $\alpha_i$
\begin{equation*}
C=\left(\begin{array}{cccc}
1 & \alpha_{2} & 0 & -\alpha_{3} \\
0 & \alpha_{1} & 1 & \alpha_{4}
\end{array}\right)
\end{equation*}

For $n=6,k=3$ we need co-dimension one boundary of the $G_+(3,6)$, which can be achieved  by setting a three minor $(i, i+1, i+2)$ to 0. We denote the 6 co-dimension one cell as $(i)$. For example the matrix corresponding to $(123)=0$ can be parametrized with $\alpha_i,i=1,\dots, 8$
\begin{equation}
    C^{(1)}=\left(
\begin{array}{cccccc}
 1 & \alpha _4+\alpha _6+\alpha _8 & \left(\alpha _4+\alpha _6\right) \alpha _7 & \alpha _4 \alpha _5 & 0 & 0 \\
 0 & 1 & \alpha _7 & \alpha _2+\alpha _5 & \alpha _2 \alpha _3 & 0 \\
 0 & 0 & 0 & 1 & \alpha _3 & \alpha _1 \\
\end{array}
\right)
\end{equation}

\subsection{ABJM}
The ABJM amplitudes also have integral representation over the orthogonal Grassmannian~\cite{Huang:2013owa,Huang:2014xza},
\begin{equation}\label{eq:og}
	\int\frac{d^{k\times2k}C_{\alpha i}}{\mathrm{Vol}\left(\mathrm{GL}\left(k\right)\right)}\frac{1}{M_{j}M_{j+1}\cdots M_{j+k-1}}\delta^{k\left(k+1\right)/2}\left(C\cdot\Omega\cdot C^{T}\right)\prod_{a=1}^{k}\delta^{2|3}\left(C_{a}\cdot\Lambda\right)
\end{equation} 
where $\Lambda=\left(\left|i\right\rangle ,\eta_{i}^{A}\right), A=1,2,3$ and $M_l$ represent the $l$-th consecutive minor:
\begin{equation}
	M_{l}\equiv\left(l\,l+1\cdots l+k\right)
\end{equation}
and the metric is $\Omega=(+,-,+,-,\cdots)$. For $k=even$, $j=1$ if $\bar{\Psi}$ is on odd sites, while $j=2$ if otherwise. For $k=odd$, $j=2$ if $\bar{\Psi}$ is on odd sites, while $j=1$ if otherwise. The choice of alternated signature  plays an important role in defining positivity of orthogonal Grassmannian and it makes the minors of the matrix $C$ be real under the orthogonal condition.

The orthogonal condition $C\cdot\Omega \cdot C^T=0$ will create two branches OG$^{\text{pos}}$ and OG$^{\text{neg}}$. OG$^{\text{pos}}$ is defined by $M_I/M_{\bar{I}}=1$ while OG$^{\text{neg}}$ is defined by $M_I/M_{\bar{I}}=-1$. Here $M_I$ means minor $(i_1 i_2\ldots)$ and $\bar{I}$ is the conjugate column of $I$. For $n=4$, only the positive branch will be allowed on the support of $C\cdot\Lambda=0$; while for $n>4$, there is a solution for each of the two branches. The full amplitudes should be identified as the sum of residue of this integral in two branches.

To localize the integral~\eqref{eq:og}, the cells need to be $n-3$ dimension (after solve the orthogonal condition). On the support of the BCFW cells, the integral comes to the canonical volume form (integral over for $OG_k$):
\begin{equation}
	\int \mathcal{J}\times \prod_{i=1}^{n-3}d\log\text{tan}\,\theta_i \, \delta^{(2|3)}(C^{\text{BCFW}}\cdot\Lambda)
\end{equation}
For the 4-point and 6-point, there is no triangulation of the cells (while for $n>6$, cells start being triangulated). The BCFW cells for $n=4,6$ are the top cells and can be parameterized as follow
\begin{equation}
	\text{OG}_4=\left(\begin{array}{cccc}
c & 1 & s & 0\\
-s & 0 & c & 1
\end{array}\right)_{n=4},\
	\text{OG}_6=\left(\begin{array}{cccccc}
\frac{s_{1}+s_{2}s_{3}}{1+s_{1}s_{2}s_{3}} & 1 & \frac{c_{1}c_{2}}{1+s_{1}s_{2}s_{3}} & 0 & -\frac{c_{1}c_{3}s_{2}}{1+s_{1}s_{2}s_{3}} & 0\\
-\frac{c_{1}c_{2}s_{3}}{1+s_{1}s_{2}s_{3}} & 0 & \frac{s_{2}+s_{1}s_{3}}{1+s_{1}s_{2}s_{3}} & 1 & \frac{c_{2}c_{3}}{1+s_{1}s_{2}s_{3}} & 0\\
\frac{c_{1}c_{3}}{1+s_{1}s_{2}s_{3}} & 0 & -\frac{c_{2}c_{3}s_{1}}{1+s_{1}s_{2}s_{3}} & 0 & \frac{s_{3}+s_{1}s_{2}}{1+s_{1}s_{2}s_{3}} & 1
\end{array}\right)
\end{equation}
here the variables $s$ and $c$ denote $\sin\theta$ and $\cos\theta$. The canonical volume form for $n=4,6$ are
\begin{equation}
	\bold{\Omega}_4=d\log\text{tan}\,\theta_i, \quad \bold{\Omega}_6=(1+s_1 s_2 s_3) \prod_{i=1}^{3}d\log\text{tan}\,\theta_i 
\end{equation}
There is the non-trivial Jacobian factor $\mathcal{J}_6=1+s_1s_2s_3$.

The more relevant to us is to consider the reduced SUSY amplitude as we discuss in the main text \eqref{eq:rudsusy}. Here, we show the explicit form of the amplitude in $n=4,6$:

\begin{equation}
    A_4=\frac{\langle 13\rangle }{\langle 12 \rangle \langle 23 \rangle } \delta^3(P) \delta^{(4)}(Q)
\end{equation}

\begin{equation}\label{eq:6pmAmp}
\begin{split}
	A_6=\delta^{3}\left(P\right)\delta^{\left(4\right)}\left(q\right)\left[ \frac{p_{135}^{2}}{c^+_{25}c^+_{41}c^+_{63}}\delta^{\left(2\right)}\left(\sum_{\bar{p},\bar{q},\bar{r}=2,4,6}\left\langle \bar{p}\bar{q}\right\rangle \eta_{\bar{r}}+\sum_{p,q,r=1,3,5}\left\langle pq\right\rangle \eta_{r}\right)\right.\\
	+\left. \frac{p_{135}^{2}}{c^-_{25}c^-_{41}c^-_{63}}\delta^{\left(2\right)}\left(\sum_{\bar{p},\bar{q},\bar{r}=2,4,6}\left\langle \bar{p}\bar{q}\right\rangle \eta_{\bar{r}}-\sum_{p,q,r=1,3,5}\left\langle pq\right\rangle \eta_{r}\right)\right]
\end{split}
\end{equation}
where
\begin{equation}
    \begin{split}
    	c^{\pm}_{\bar{r}s}=\left\langle \bar{r}\right|P_{135}\left|s\right\rangle \pm \left\langle \bar{r}-2,\bar{r}+2\right\rangle \left\langle s-2,s+2\right\rangle 
    \end{split}
\end{equation}

\section{Proof of the relation $\langle Yij\rangle= g  \langle ij\rangle$}\label{sec:bracket}

Here we prove that brackets in $(Y, \tilde{Y})$ space and those in kinematic space are simply related (with constant $g$ defined in~\cite{Damgaard:2019ztj}): 
\begin{equation} 
\begin{split}
\left\langle i\,j\right\rangle & = \sum_{\alpha,\beta=1,2}\epsilon_{\alpha\beta}\left(Y^{\perp}\cdot\Lambda_{i}\right)^{\alpha}\left(Y^{\perp}\cdot\Lambda_{j}\right)^{\beta} \\
& = \sum_{\alpha,\beta=1,2}\sum_{p,q=1}^{k{+}2}\epsilon_{\alpha\beta}\left(Y_{\alpha,p}^{\perp}\cdot\Lambda_{p,i}\right)\left(Y_{\beta,q}^{\perp}\cdot\Lambda_{q,j}\right)\\
&=\sum_{p,q=1}^{k{+}2}\left(p\,q\right)_{Y^{\perp}}\Lambda_{p,i}\Lambda_{q,j}\\
	&=\sum_{p,q=1}^{k{+}2}g\epsilon_{pqr_{1}\cdots r_{k}}\left(r_{1}\cdots r_{k}\right)_{Y}\Lambda_{p,i}\Lambda_{q,j}
	\end{split}
\end{equation}
where $\left(p\,q\right)_{Y^{\perp}}=\epsilon_{\alpha\beta}Y_{\alpha,p}^{\perp}Y_{\beta,q}^{\perp}$, which can be rewritten as \\$$g\,\epsilon_{pqr_{1}\cdots r_{k}}\left(r_{1}\cdots r_{k}\right)_{Y}=\sum_{l_{i}=1}^{k}g\,\epsilon_{pqr_{1}\cdots r_{k}}\epsilon_{l_{1}l_{2}\cdots l_{k}}Y_{l_{1},r_{1}}Y_{l_{2},r_{2}}\cdots Y_{l_{k},r_{k}}$$ and $\left\{ r_{1},\cdots,r_{k}\right\} =\left\{ 1,2,\cdots\hat{p},\cdots\hat{q},\cdots k{+}2\right\}$. Then, we can use the identity, 

\begin{equation}
	\begin{split}
		& \sum_{p,q=1}^{k{+}2}\sum_{l_{i}=1}^{k}g\epsilon_{pqr_{1}\cdots r_{k}}\epsilon_{l_{1}l_{2}\cdots l_{k}}Y_{l_{1},r_{1}}Y_{l_{2},r_{2}}\cdots Y_{l_{k},r_{k}}\Lambda_{p,i}\Lambda_{q,j}\\
		=&\sum_{p,q=1}^{k{+}2}\sum_{r_{i}=1}^{k{+}2}g\epsilon_{pqr_{1}\cdots r_{k}}Y_{1,r_{1}}Y_{2,r_{2}}\cdots Y_{k,r_{k}}\Lambda_{p,i}\Lambda_{q,j}\\=& g \left\langle Yij\right\rangle,
	\end{split}
\end{equation}
and similarly for $[i j]=\tilde{g} [ \tilde{Y} i j]$. This completes the proof.

\bibliographystyle{utphys}
\bibliography{bib}

\end{document}